\documentclass[manuscript]{acmart}
\AtBeginDocument{%
  \providecommand\BibTeX{{%
    \normalfont B\kern-0.5em{\scshape i\kern-0.25em b}\kern-0.8em\TeX}}}

\setcopyright{acmcopyright}
\copyrightyear{2025}
\acmYear{2025}
\acmDOI{XXXXXXX.XXXXXXX}

\acmConference[CHI '25]{ACM Conference on Human Computer }{Date}{Location}
\usepackage{caption}
\usepackage{subcaption}
\usepackage{multirow}
\usepackage{alphabeta}
\usepackage{booktabs}
\usepackage{array}
\usepackage{tabularx}
\usepackage{adjustbox}
\usepackage{makecell}
\usepackage{float}
\usepackage{xcolor}
\usepackage{ulem}

\author{Dashiel Carrera}
\email{dcarrera@dgp.toronto.edu}
\orcid{0000-0001-9851-3733}
\affiliation{%
  \institution{University of Toronto}
  \city{Toronto}
  \state{Ontario}
  \country{Canada}
}

\author{Jeb Thomas-Mitchell}
\email{jeb.thomasmitchell@mail.utoronto.ca}
\orcid{0009-0000-7403-2329}
\affiliation{%
  \institution{University of Toronto}
  \city{Toronto}
  \state{Ontario}
  \country{Canada}
}

\author{Daniel Wigdor}
\email{daniel@dgp.toronto.edu}
\orcid{0000-0003-2008-7070}
\affiliation{%
  \institution{University of Toronto}
  \city{Toronto}
  \state{Ontario}
  \country{Canada}
}

\begin{document}

\title{Where Do I `Add the Egg?': Exploring Agency and Ownership in AI Creative Co-Writing Systems} 




\begin{abstract}
AI co-writing systems challenge long held ideals about agency and ownership in the creative process, thereby hindering widespread adoption. In order to address this, we investigate conceptions of agency and ownership in AI creative co-writing. Drawing on insights from a review of commercial systems, we developed three co-writing systems with identical functionality but distinct interface metaphors: agentic, tool-like, and magical. Through interviews with professional and non-professional writers (n = 18), we explored how these metaphors influenced participants’ sense of control and authorship. Our analysis resulted in a taxonomy of agency and ownership subtypes and underscore how tool-like metaphors shift writers’ expected points of control while agentic metaphors foreground conceptual contributions. We argue that interface metaphors not only guide expectations of control but also frame conceptions of authorship. We conclude with recommendations for the design of AI co-writing systems, emphasizing how metaphor shapes user experience and creative practice.
\end{abstract}

\maketitle

\section{Introduction}
The advent of Large Language Models (LLMs) has opened new artistic possibilities for creative writers, resulting in novel forms of poetry, generative novels, and computational literature \cite{whalen2023any, marche2023death}. This relatively new yet increasingly pervasive form of Artificial Intelligence (AI) has lowered the barrier to entry for amateur or unpracticed creative writers, enabling them to explore genres and forms that may have previously felt inaccessible. Despite the new artistic possibilities brought about by AI, writers' perceptions of these platforms remain mixed, with many skeptical of whether AI can yield sufficiently creative results \cite{chakrabarty2024artoratifice} or whether the introduction of AI into their writing practice would disrupt or threaten their creative integrity \cite{cnnTheseBooks, wasi2024ink}. 

One of the ways writers perceive their interactions with AI co-writing systems is through their perception of \textit{agency} when they use the system and their perception of \textit{ownership} over the final writing product \cite{Gero23}. A writer's perception of agency is the extent to which they feel a sense of control while using the system, and a writer's perception of ownership is the extent to which they self-identify as an author of the writing co-produced with the AI. To support broader adoption of AI co-writing systems among writers, it is crucial to consider perceived agency and perceived ownership, as these factors can influence a writer's long-term satisfaction with such systems~\cite{Gero23}. 

Previous work has identified agency and ownership as key factors in how writers experience AI co-writing \cite{Lee24} and has characterized what types of ownership matter to writers in general \cite{reza2025co}. Additional work has demonstrated that agency and ownership are generally correlated: writers who feel greater control while AI co-writing generally feel greater ownership over the results \cite{Draxler24}. However, little is known about how creative writers conceptualize agency and ownership during interactions with AI creative co-writing systems. This is especially surprising due to creative writing's rich history of deriving multiple conceptualizations of agency and ownership in the creative process: from listening to the muse \cite{shelley1910defence} to post-modern challenges of authorship \cite{barthes2016death} to aleatory or probabilistic methods of creation \cite{tzara2018seven}. 

To explore agency and ownership during AI creative co-writing, we use interface metaphors as a design probe. 
An interface metaphor is a design concept where familiar objects or concepts are used to represent and facilitate interaction with digital systems, making them easier to understand \cite{marcus1992graphic}. 
Phil Agre's Critical Technical Practice \cite{agre1997computation} argues that by analyzing the particular metaphors and language of a field of technology, we can identify what the field marginalizes. Interface metaphors are one of the most effective means of altering user perceptions and mental models \cite{Carroll1988Interface}, having previously been used in mobile interfaces and voice-based user interfaces \cite{Li2022Navigating, Desai24}. We seek to understand how interface metaphors marginalize or afford particular conceptions of agency and ownership. Additionally, by presenting participants with different metaphors, we elicit a broader range of conceptualizations of agency and ownership than those currently available in the literature. In furtherance of this exploration, we introduce the following research questions: 

\begin{quote}
\textbf{RQ1}: What are the different ways writers conceptualize \textbf{agency} and \textbf{ownership} in AI creative co-writing? 

\textbf{RQ2:} How does the use of \textbf{interface metaphors} in AI creative co-writing systems affect these conceptualizations? 
\end{quote}

In order to determine which interface metaphors to study, we performed a review of popular interface metaphors used in commercial AI creative co-writing systems today. Based on this review and prior literature \cite{Lee24}, we identified three interface metaphor archetypes that were the most prevalent: tool-like, agentic, and magic. We then created three AI co-writing systems, each with a different interface features that exemplified each metaphor. 

We conducted an exploratory study followed with a semi-structured interview (n = 18) where nine professional and nine non-professional writers with creative writing experience were presented with one of three interface metaphors. In the interview, we explored how writers conceptualized agency and ownership in their co-writing experiences. Our analysis revealed multiple forms of agency and control in AI creative co-writing, including three subtypes of control—explicit, implicit, and process-based—and three subtypes of ownership—stylistic, conceptual, and effort-based. Notably, agentic metaphors fostered conceptions of ownership that emphasized the AI’s conceptual contributions.

This paper makes three contributions. First, it presents a preliminary review of interface metaphors in common commercial AI co-writing systems, which should ground future work on AI co-writing system interfaces. Second, it introduces a taxonomy of agency and ownership in AI creative co-writing. Finally, it offers design recommendations for future AI co-writing systems, highlighting how interface metaphors shape writers’ expectations of control and authorship.
 
\section{Related Work}
In this section, we detail research on AI co-writing systems that describes the common dynamics and expectations writers have for these systems and examine conceptions of agency and ownership for users of these systems. We also review research that describes how interface metaphors shape a user's expectations and interactions with an AI system.

\subsection{AI Writing systems}
The idea that computational processes could contribute to writing has a long history, with a very early example being mechanical machines that could generate lines of Latin verse \cite{sharples2022story}. Early work in Computer Science, by Roemmele, presented an AI co-writing system for writers that generated suggestions for the next sentences in a story using information retrieval techniques \cite{roemmele_creative_2015}. Later work by Roemmele and Gordon extended this application to use neural networks \cite{roemmele_writing_2016}. A 2024 literature review of intelligent and interactive writing assistants by Lee et al. found that since the mid-2010s, the number of papers published on writing assistants has increased dramatically \cite{Lee24}. Looking at creative writing specifically, there has been research on new techniques to support story writing \cite{yuanWordcraftStoryWriting2022, ippolitoWordCraftCreativeWritingAIPowered2022}, playwriting \cite{Mirowski23}, character development \cite{qinCharacterMeetSupportingCreative2024}, prewriting \cite{wan2024ItFeltHaving}, nonlinear story writing (also known as 'multiverse') \cite{Ghajargar22}, worldbuilding \cite{chung2024PatchviewLLMPoweredWorldbuilding}, cartoon-captioning \cite{kariyawasam2024AppropriateIncongruityDriven}, and feedback generation \cite{benharrakWriterDefinedAIPersonas2024}. Such research has demonstrated that AI has the potential to support a variety of creative writing activities and aspects of the creative writing process. However, such research remains in its early stages and is largely demonstrative, as these systems have yet to see widespread adoption among writers and are typically studied in small, lab-based settings.

In addition to systems that support specific creative writing activities, there has also been work investigating how creative writers interact with, and reason about, writing support. CoAuthor provided a dataset for studying AI co-writing and demonstrated how tracing writing interactions can allow for rigorous investigations into human-computer collaboration \cite{Lee22}. Other projects have investigated how creative writers incorporate AI suggestions, for instance, engaging in 'integrative leaps' \cite{singhWhereHideStolen2022} or viewing suggestions antagonistically \cite{calderwoodHowNovelistsUse2018}. Gero et al. reported on how writers reason about AI support differently than human support \cite{Gero23} and Kim et al. identified the benefits of AI co-writing based on the dynamics between an author, artifact, and the audience \cite{kim2024AuthorsValuesAttitudes}. Such work has also outlined several issues that arise during AI co-writing, including a decreased sense of agency and ownership, which we consider next.



\subsection{Agency and Ownership with AI Co-Writing}
Early work on understanding AI co-writing in creative contexts suggested that AI involvement may decrease writers’ sense of ownership \cite{geroMetaphoriaAlgorithmicCompanion2019, geroSparksInspirationScience2022}. More recent work on AI co-writing for creative and argumentative texts has explicitly investigated issues of agency and ownership, finding that the less a user writes, the less ownership they feel over the final text \cite{Lee22}. Others have found that AI assistance for idea generation decreases feelings of ownership \cite{guo2024ExploringImpactAI} and that suggesting whole paragraphs of text may increase writing quality but at the cost of decreased feelings of ownership \cite{dhillon2024ShapingHumanAICollaboration}. 

Other work has investigated additional factors that impact perceptions of ownership. For example, Kadoma et al. found that the style of AI suggested text (e.g., hesitant or assertive) can impact writers’ perceptions of ownership \cite{kadoma2024RoleInclusionControl}. In a two-week diary study, Kobiella et al. investigated young professionals' perceptions of ChatGPT and found that feelings of ownership were related to how much they felt their contribution was central to the task at hand \cite{kobiella2024IfMachineGood}. 
While investigating the relationship between authorship and ownership, Draxler et al. found that higher levels of participant influence on text increased one's sense of ownership and that ownership and authorship were not always neatly correlated \cite{Draxler24}. Participants were more likely to claim authorship of AI-assisted texts than human-assisted ones. Reza et al. recently identified a distinction between academic and creative writers' values in ownership, finding that academic writers valued the ideas and content of pieces they created, while creative writers valued the formal and linguistic elements \cite{reza2025co}. Hoque et al. designed a system to encourage agency and ownership that tracked the provenance of text and found that provenance information increased a writer’s sense of agency and ownership by supporting active considerations of what the AI contributed \cite{hoque2024HaLLMarkEffectSupporting}. Such work demonstrates that perceptions of ownership are not only a function of how much text was generated by an AI assistant but were also influenced by perceptions of the AI as a collaborative partner and judgments about its contribution.


\subsection{Interface Metaphors for AI}
Users may have preconceived notions about AI systems, so metaphors can help users understand the capabilities of an AI agent by proposing an alternative mental model \cite{Gero23, ChinDesai24}. The mental models evoked by a metaphorical comparison facilitate users' familiarization with a system by creating a set of expectations around its functionality \cite{Desai23}. By incorporating different metaphors into AI systems, designers can create shared meanings that help users express nuanced views about, and experiences with, an AI agent \cite{Desai23}. 

Comparative research on metaphors for AI has demonstrated the impact metaphors can have on user interactions with task-based AI systems, e.g., AI that assists with well-defined tasks such as travel planning \cite{Khadpe20} or waste sorting \cite{Papachristos21}. For example, Khadpe et al. \cite{Khadpe20} explored agentic metaphors that portrayed chatbots as different human personas and found that they changed user satisfaction with a system even when the system's capabilities were unchanged.  Users have also been found to project their own metaphors onto AI, which influenced their interactions \cite{Papachristos21, ChinDesai24}. Chin and Desai investigated older adults' use of voice assistants and found that participants ascribed a variety of object- and human-based metaphors that suggested different conceptions of the assistant's agency, autonomy, and animacy \cite{ChinDesai24}. This comparative research has shown that interface metaphors, whether explicitly assigned by designers or implicitly adopted by users, significantly influence users' perceptions and interactions with AI systems.

In the domain of AI co-writing systems, interface metaphors have also been found to impact the types of interactions that users expect to encounter. Based on a review of prior literature, Lee et al. described three \textit{interaction metaphors} for AI co-writing systems. "Agent" metaphors described designs that used anthropomorphized AI, with conversational interactions and streaming text that simulated an interaction with another person. "Tool" metaphors compartmentalized an AI system’s capabilities and presented controls using familiar interface elements like buttons and dropdown menus. Systems designed with "hybrid" metaphors mixed elements from "agents" and "tools" \cite{Lee24}. Interaction metaphors, as used in Lee et al., are closely related to interface metaphors, and they are often deployed in concert to present intuitive affordances to users. While interface metaphors relate visual and organizational design to other objects through metaphor, interaction metaphors relate the actions a user can perform. Lee et al.'s use includes both interactions (e.g. conversational interactions) and interface presentation (e.g. streaming text), so their findings are relevant to our study of interface metaphors. Biermann et al. used a design workbook approach to compare different AI interfaces and investigate how writers' personal values affected their intent to adopt AI in their practices \cite{Biermann22}. Their study revealed differences in the preferred division of control between the writer and AI based on the interface presented and the writer's writing process, professional status, feelings of self-efficacy, and emotional attachment to their prose. Our research extends this prior work by using the demonstrated perceptual impacts of AI interface metaphors to modulate and examine how writers conceptualize agency and ownership in their interactions with AI co-writing systems.

\section{Interface Metaphors in Commercial AI Creative Co-Writing Systems}


To facilitate the design of interface metaphor systems that would closely mimic the conditions that writers are most likely to encounter, we conducted a review of the interface metaphors that were most often used within commercial AI co-writing systems. As different metaphorical forces can have a combined effect on the interaction process, identifying and later isolating each metaphor under controlled conditions enables us to develop a more detailed understanding of how these metaphors may influence users of commercial interfaces. Additionally, by identifying archetypal metaphors from which hybrid metaphors are most commonly composed, we can (1) map out the design space of interface metaphors in AI creative co-writing systems and (2) explore a range of different interaction dynamics in AI creative co-writing that elicit a broad range of different conceptions of agency and ownership. 

Thus, in May 2024, we conducted a review of existing commercial AI co-writing systems was based on prior HCI research that compared commercial systems \cite{Bentvelzen22}. While academic research has explored various interface approaches for AI writing assistance, we focused on a review of commercial systems because they include the interface metaphors that writers are most likely to encounter, and because these systems could be readily and thoroughly tested by the research team. 

Our review began with a preliminary search for popular AI co-writing systems using Google and the Reddit writing communities r/writingwithai\footnote{https://www.reddit.com/r/WritingWithAI/} and r/writing\footnote{https://www.reddit.com/r/writing/}. This search resulted in 45 AI co-writing systems. We then categorized the systems based on their intended audience and use cases, using their marketing materials as reference. We excluded systems that were not targeted at creative writers, which left a total of nine systems. We also included ChatGPT, Google Gemini, and Grammarly due to their popularity and prevalence within the aforementioned creative writing communities. This led to a total of 12 AI creative co-writing systems that were reviewed: \href{https://chatgpt.com/}{ChatGPT}, 
\href{https://dreamgen.com/}{DreamGen}, 
\href{https://fictionary.co/}{Fictionary}, 
\href{https://docs.google.com/}{Google Gemini in Google Docs}, 
\href{https://www.grammarly.com/}{Grammarly}, 
\href{https://novelai.net/}{NovelAI}, 
\href{https://www.novelcraft.net/}{NovelCraft}, 
\href{https://www.novelcrafter.com/}{Novelcrafter}, 
\href{https://quillbot.com/}{Quillbot}, 
\href{https://www.squibler.io/}{Squibler}, 
\href{https://www.sudowrite.com/}{Sudowrite}, 
and \href{https://writesonic.com/}{Writesonic}.

Our analysis included an hour-long exploration of all the features and interface elements in each system. We captured screen recordings of each system and recorded the textual metaphors presented in text descriptions and iconographic metaphors in icons representing AI assistance. We then characterized each system based on the interface presentation they used (e.g., icons and descriptions), the interaction models they afforded to the user, and the user agency they enabled or constrained. We began our review using the framework from Lee et al., \cite{Lee24} who characterized "Tool" and "Agent" interface metaphors, along with hybrid approaches. We found clear correspondences between their descriptions for "Tool" and "Agent" and the metaphorical presentations we observed in many systems. While some systems \cite{sudowrite, quillbot} did present a mix of metaphors, we found that most closely reflected a single interface metaphor, so we chose not to include the hybrid metaphors as an archetype. We also identified magic-like metaphors in nine systems, which had distinctive interface presentations and interaction designs. Our review revealed three recurring archetypes of how AI interactions were presented to users through interface elements: Agent-like, Tool-like, and Magic-like (Table \ref{tab:interface-archetypes}). While the archetypes are not an exhaustive categorization of all possible AI interface metaphors, they emerged as the most prevalent approaches found in current commercial creative co-writing systems. Summary statistics of the review can be found in the Appendix \ref{CASST} (Table \ref{fig:designreference}). 

\begin{table}[h]
\centering
\caption{The three archetypal interface metaphors found in AI co-writing systems.}
\label{tab:interface-archetypes}
\small 
\begin{tabular}{>{\raggedright\arraybackslash}p{1.8cm}>{\raggedright\arraybackslash}p{2.8cm}>{\raggedright\arraybackslash}p{2.5cm}>{\raggedright\arraybackslash}p{2.3cm}>{\raggedright\arraybackslash}p{2.2cm}>{\raggedright\arraybackslash}p{1.8cm}}
\toprule
\textbf{Archetype} & \textbf{Characteristics} & \textbf{Interface Elements} & \textbf{AI Interaction} & \textbf{User Agency} & \textbf{Examples} \\
\midrule
\textbf{Agent-like} & Personified AI; collaborative partnership; natural language interface & Chat windows, streaming text, persona icons, first-person pronouns & Conversational; iterative; often separated from main text area & Shared agency; AI as collaborative partner & ChatGPT, Novelcrafter (Nia), DreamGen \\
\midrule
\textbf{Tool-like} & Precise control over AI capabilities; compartmentalized functions; mechanical imagery & Buttons, drop-down menus, sliders, keyword inputs for length, tone, and style & Direct manipulation; integrated with text editing & Fine-grained user control; AI as instrument extending user's creative agency & Squibler (Smart Write), Writesonic (Continue Writing) \\
\midrule
\textbf{Magic-like} & Simple automatic functions; magical iconography; emphasis on seamlessness & Wands, sparkles, minimal controls, one-click generation & One-shot automation; minimal iteration & Minimal explicit control; automatic "enchanted" generations & Sudowrite (Auto Write), NovelCraft \\
\bottomrule
\end{tabular}
\end{table}

\textbf{Agent-like interfaces}, found in seven systems, were primarily implemented through conversational interfaces with a personified AI that could accomplish a variety of functions. Some systems enhanced this sense of agency through specific personas applied to the AI, such as Novelcrafter's support AI Nia, which took the form of a cute goblin \cite{novelcrafter}. These interfaces emphasized natural language interaction and often presented the AI as a collaborative partner that addressed the user and directly contributed in the writing area. This archetype was typified by ChatGPT (Figure \ref{fig:chatgpt}), which featured a chat interface that streamed text to mimic typing by the agent and framed the AI as a personified agent through its use of first-person pronouns and directly addressing the user. DreamGen \cite{dreamgen} employed similar streaming text pattern to reinforce its agent-like presence.

\textbf{Tool-like interfaces}, present in 10 of the 12 systems, offered precise control over AI capabilities and compartmentalized functions to diminish a user's perception of independent AI agency. These elements featured mechanical imagery such as pencils or gears, and provided explicit interface elements for users to constrain the length, tone, and style of AI-generated content (e.g., buttons, dropdown menus, sliders, and keywords). For example, Squibler's Smart Write interface (Figure \ref{fig:squibler}) provided explicit control elements such as sliders for the amount of dialogue in the text, a text box to define the content of the generated section, and a dropdown menu to select the narrative perspective. Writesonic's Continue Writing feature \cite{writesonic} similarly exemplified this approach through its interface's parameter controls.

\textbf{Magic-like interfaces}, present in nine systems, did not fall along the Tool-Hybrid-Agent spectrum described by Lee et al. \cite{Lee24}. These interfaces offered simple and automatic functions, and used magical iconography such as wands and sparkles. Similar to how Tool-like interfaces offered compartmentalized functionality, these interface elements were distinguished by their lack of precise control over the AI output. Sudowrite's Auto Write feature (Figure \ref{fig:sudowrite}) emphasized automaticity and limited user controls compared to other tools, but had textual and iconographic references to magic. NovelCraft \cite{novelcraft} reinforced magical framing through wand iconography, while sparkles (which have been found to contribute to the perception of magic \cite{Hui15}) appeared across eight of the 12 systems. The application of magic metaphors to AI has been previously explored in HCI and Design research \cite{Lupetti24}, and the application of magic metaphors to computing devices has a long history in HCI \cite{Svanaes00, Guldenpfennig19, Marshall10, Ciger03}.

\section{Exploratory Interview-Based Study of Interface Metaphors for AI Co-Design Interfaces}


We conducted an exploratory study with semi-structured interviews with professional and non-professional creative writers to understand their conceptions of aspects of agency (e.g., ease and precision of control over the system, control over the writing process and whether the system supported and aligned with the writer’s goals) and ownership (e.g., authorship, collaborative authorship, representation of the writer’s unique style and voice, etc.) when using AI co-writing systems. Participants first completed a short creative writing task using a prototype framed with one of three interface metaphors, followed by an interview with two researchers. During the interviews, participants were asked how their collaboration with the AI could change if the generated text was less similar to their voice or style, if their interaction surprised them, and if they felt like the AI was a co-author. The study was approved by our institution's Research Ethics Board and each participant provided written and verbal consent before beginning the study. 

\subsection{Participants}
We recruited 18 participants using the first author's network of fellow literary writers, social media, and outreach within our institution. We targeted two distinct groups: professional creative writers (n = 9) and non-professional writers (n = 9). Eleven participants self-identified as men and 7 participants self-identified as women. 

Our study recruited creative writers from the North American English-language literary fiction community, a community which endeavors to create works of artistic merit and cultural importance. In English fiction, such writers generally emphasize precision of language, style, and nuance of character. We chose to study these writers because their interest in precise style and deep consideration for various modes of creative agency and authorship makes them distinct from the general population. The non-professional writer group was included to provide a contrasting perspective and to explore how non-professional writers might perceive agency and ownership in AI co-writing. System use and the interview together took between 60 and 90 minutes per participant. Participants received a \$45 CAD Amazon gift card as an honorarium for their participation.

\begin{table}
\centering
\small
\begin{tabular}{| l | l | l | l | l |}
\hline
\textbf{Participant} & \textbf{Gender} & \textbf{Occupation} & \textbf{Group} & \textbf{AI Interface} \\
\hline
1 & M & Novelist, Playwright & Professional Creative Writer & Autowrite (Tool-like) \\
\hline
2 & M & Novelist, Short Story Author & Professional Creative Writer & Autowrite (Tool-like)  \\
\hline
3 & M   & Poet, Short Story Author & Professional Creative Writer& Wordsworth (Agent-like) \\
\hline
4 & M  & Novelist, English Professor & Professional Creative Writer & Magic Quill (Magic-like)\\
\hline
5 & M & Novelist, Short Story Author & Professional Creative Writer & Magic Quill (Magic-like)\\
\hline
6 & F & AI Consultant & Non-Professional Writer & Autowrite (Tool-like)\\
\hline
7 & F  & PR Consultant & Non-Professional Writer & Magic Quill (Magic-like)\\
\hline
8 & F & Short Story Author & Professional Creative Writer  & Magic Quill (Magic-like)\\
\hline
9 & M & Comedy Writer & Professional Creative Writer & Wordsworth (Agent-like) \\
\hline
10 & F & HCI Researcher & Non-Professional Writer & Magic Quill (Magic-like)\\
\hline
11 & F  & Animation Researcher & Non-Professional Writer & Wordsworth (Agent-like)\\
\hline
12 & F & Film Producer & Non-Professional Writer & Magic Quill (Magic-like)\\
\hline
13 & M  & Engineering Student & Non-Professional Writer & Wordsworth (Agent-like) \\
\hline
14 & M & Poet, Short Story Author & Professional Creative Writer & Autowrite (Tool-like) \\
\hline
15 & M  & Communications Student & Non-Professional Writer & Wordsworth (Agent-like)\\
\hline
16 & M & Educator & Non-Professional Writer & Autowrite (Tool-like)\\
\hline
17 & M & Prop Master, Art Director & Non-Professional Writer & Autowrite (Tool-like)\\
\hline
18 & F & Novelist, Short Story Author & Professional Creative Writer & Wordsworth (Agent-like)\\
\hline

\end{tabular}
\caption{Demographics of the Exploratory Study Participants.}
\end{table}

\subsection{AI Creative Co-Writing Systems}
\label{sec:prototype_design}
\begin{figure}
\centering
\begin{subfigure}{.5\textwidth}
  \centering
  \includegraphics[width=1\linewidth]{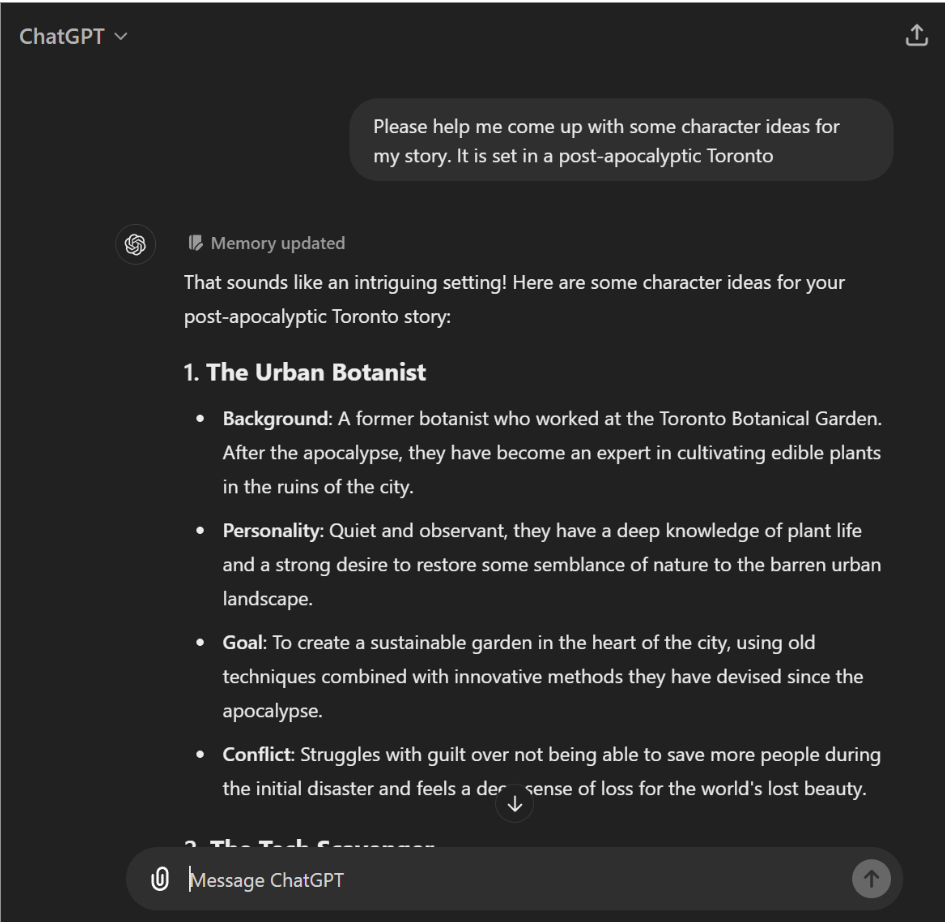}
  \caption{Agent-like Commercial Reference Example: ChatGPT}
  \label{fig:chatgpt}
\end{subfigure}%
\begin{subfigure}{.5\textwidth}
  \centering
  \includegraphics[width=1\linewidth]{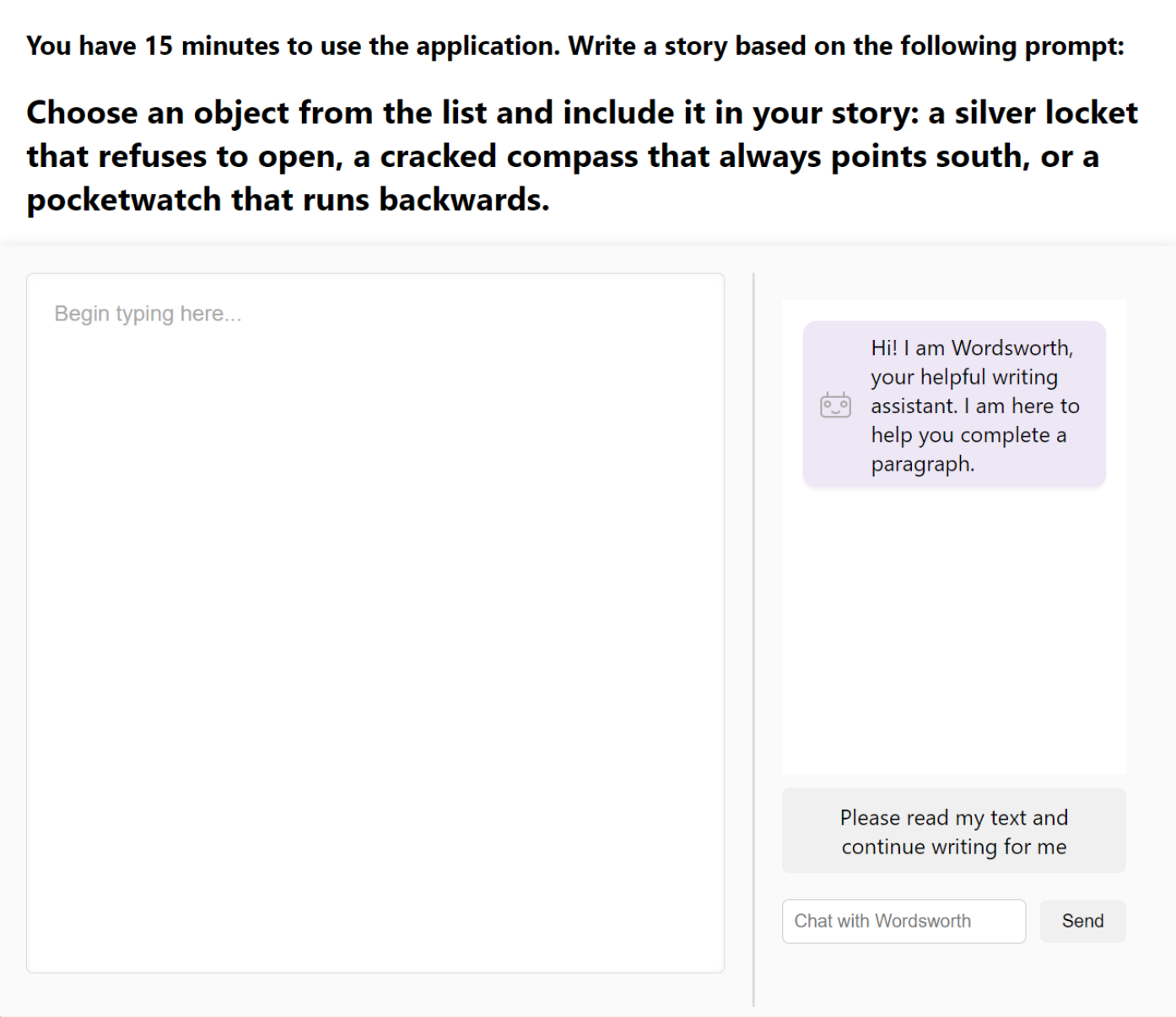}
  \caption{Wordsworth Interface}
  \label{fig:wordsworth}
\end{subfigure}
\caption{(a) A reference example of an agent-like interface metaphor: ChatGPT's conversational interface used a chat window to communicate with the user. (b) The Wordsworth interface had a chat window in the right panel for conversational interactions with the AI.}
\label{fig:wordsworth_example_reference}
\end{figure}

To investigate how interface metaphors influence writers' perceived agency and ownership in AI creative co-writing systems, we developed three prototype AI co-writing systems: Wordsworth (agentic-like), Autowrite (tool-like), and Magic Quill (magic-like). Each prototype was presented with a name that also reinforced the metaphorical framing. The prototypes were designed with near-identical functionality to minimize confounding variables. Similar to existing commercial systems, each co-writing system had a text editing area on the left where the participant could write a story. A panel on the right contained interface elements that represented one of the three AI metaphors (example pictured in Figure \ref{fig:wordsworth}).

The AI-generated text appeared in the editing area as a completion of the existing content, appended to the end. Completions were generated by GPT-3.5-turbo via OpenAI’s API and were limited to 250 tokens (approximately 100 words). Each completion was preceded by the system prompt: \textit{“Suggest a continuation for this prompt. Your reply should only be the continuation. The continuation should be grammatically correct and flow naturally from the prompt. The continuation should be paragraph-length.”} The web-based prototype used \href{https://react.dev/}{React} for the front-end, \href{https://flask.palletsprojects.com/}{Flask} for the back-end, and was hosted on \href{https://www.heroku.com/}{Heroku}.




\paragraph{Wordsworth}
Wordsworth implemented the agent-like archetype through a chat interface with a stylized robot icon and human waving icon to reinforce Human-Agent collaboration (Figure \ref{fig:wordsworth_example_reference}). The system streamed text, character-by-character, into the text area on the left to simulate typing. 

 While designed with a chat interface, the system's functionality was constrained to match non-conversational completion models. Wordsworth was limited to basic text completion and could not accommodate requests for stylistic modifications or auxiliary functions such as research. Participants were able to request completions in natural language in the chat. GPT-3.5-turbo parsed chat queries to identify completion requests, but all conversational interactions in the chat window employed scripted responses rather than AI-generated replies. To reduce interaction friction and make invocation comparable to the other systems, we also included a one-click button labeled \textit{"Please read my text and continue writing for me"} that automatically sent this request in the chat to Wordsworth.

\begin{figure}
\centering
\begin{subfigure}{.5\textwidth}
  \centering
  \includegraphics[width=1\linewidth]{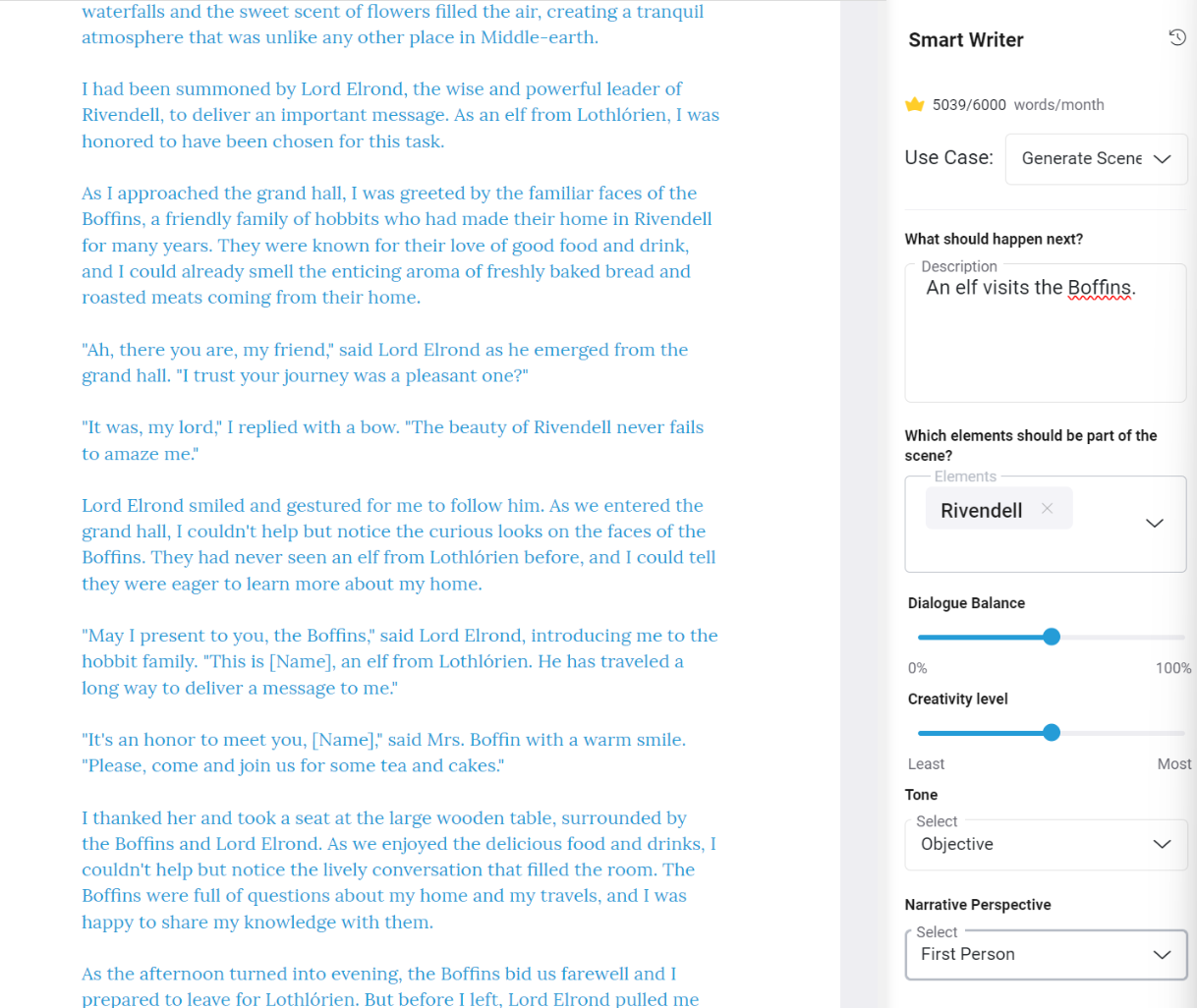}
  \caption{Tool-like Commercial Reference Example: Squibler}
  \label{fig:squibler}
\end{subfigure}%
\begin{subfigure}{.5\textwidth}
  \centering
  \includegraphics[width=1\linewidth]{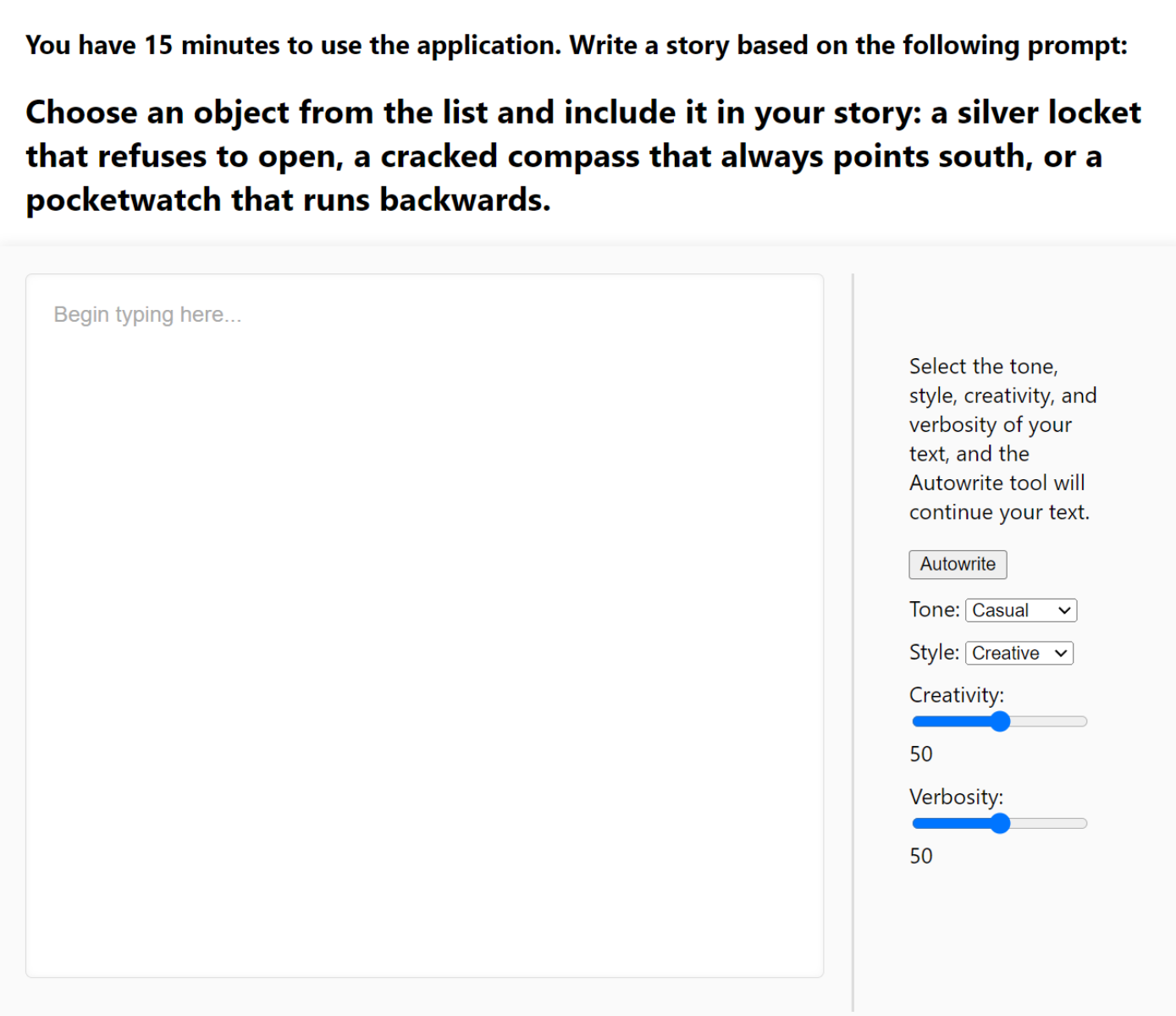}
  \caption{Autowrite Interface}
  \label{fig:autowrite}
\end{subfigure}
\caption{(a) A reference example of a tool-like interface metaphor: Squibler's Smart Writer interface used multiple control elements such as dropdown menus and sliders that offered fine-grained control over the generated text. (b) The Autowrite interface had dropdown menus and sliders in the right panel to provide the illusion of fine-grained control over the generated text.}
\label{fig:autowrite_example_reference}
\end{figure}

\paragraph{Autowrite}
Autowrite instantiated the tool-like archetype using dropdown menus for "Tone" (e.g., Casual, Formal, Academic) and "Style" (e.g., Creative, Technical, Narrative), and also had two sliders for "Creativity" and "Verbosity" that were adjustable from 0 to 100 (Figure \ref{fig:autowrite_example_reference}). Although these controls gave participants the impression of precise control, they did not actually modify the AI's output---a deception necessary to maintain consistent functionality across all three systems. Clicking the "Autowrite" button delivered the complete AI-generated text after approximately one second.

\begin{figure}
\centering
\begin{subfigure}{.5\textwidth}
  \centering
  \includegraphics[width=1\linewidth]{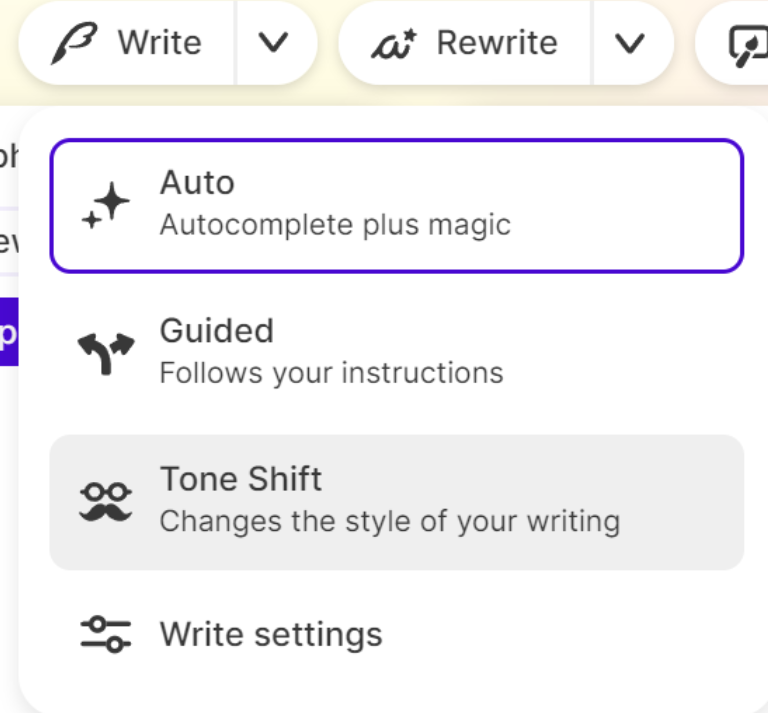}
  \caption{Magic-like Commercial Reference Example: Sudowrite}
  \label{fig:sudowrite}
\end{subfigure}%
\begin{subfigure}{.5\textwidth}
  \centering
  \includegraphics[width=1\linewidth]{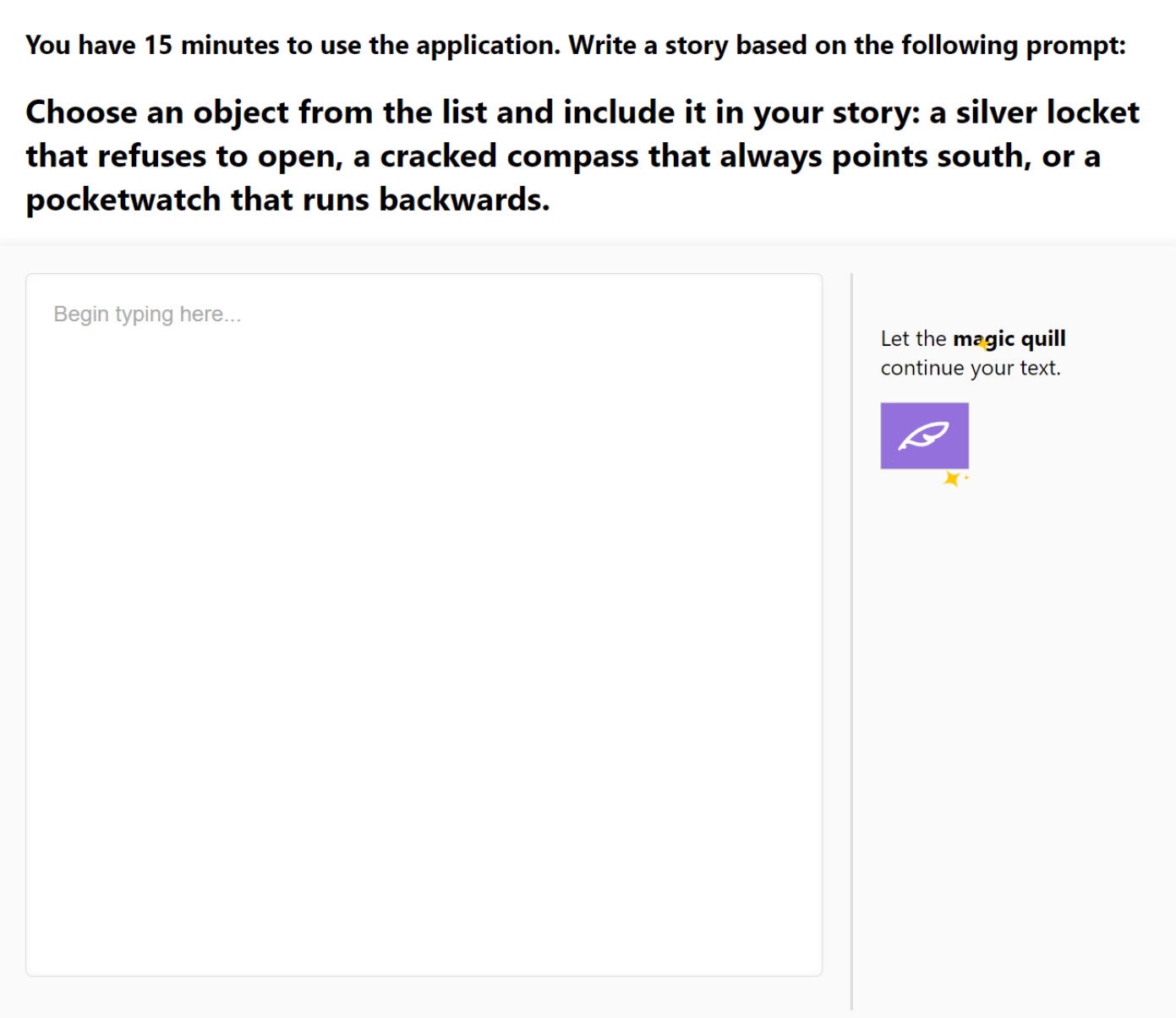}
  \caption{Magic Quill Interface}
  \label{fig:magicquill}
\end{subfigure}
\caption{(a) A reference example of a magic-like interface: Sudowrite's Write button menu used a sparkle icon, the text "Autocomplete plus magic", and withheld other interface controls to invoke magic as a metaphor for creative generation. (b) The Magic Quill interface featured a single point of control, animated sparkles around the quill button, and the text "magic quill."}
\label{fig:magic_quill_example_reference}
\end{figure}

\paragraph{Magic Quill}
Magic Quill embodied the magic-like archetype using minimal, iconographic controls: a single button featuring a quill icon with continuous gentle sparkle animations (Figure \ref{fig:magic_quill_example_reference}). When clicked, AI-generated text faded into the text area over approximately one second, accompanied by intensified sparkle animations across the interface. The fade-in effect and sparkles were designed to evoke enchanted writing from popular culture, such as Tom Riddle's diary from Harry Potter\footnote{Fade-in of enchanted text seen in this clip from Harry Potter and the Chamber of Secrets: https://www.youtube.com/watch?v=eh4b5zC0sB4}.

\subsection{Procedure}
The exploratory study was conducted online via Zoom and transcribed by \href{https://otter.ai}{Otter}. At the start of each session, researchers obtained consent to record and transcribe audio from the participant. Any concerns about data management or study procedures were addressed before proceeding. 

Introductory questions then sought to establish rapport and gather information about participants' writing processes and experiences with AI tools. These questions explored typical writing routines and familiarity with, or attitudes towards, AI co-writing systems. After the introductory questions, participants were introduced remotely to one of the three AI co-writing systems. This exploratory used a between-subjects design since the effects of the interface metaphors are confounded if the participant is exposed to more than one during their interaction. The conditions were randomly assigned while ensuring an even distribution across the professional writer and non-professional writer groups. The research team explained the system and allowed participants to familiarize themselves with its functionality. Participants then engaged with the system for 15 minutes while they were observed by the research team. While a writing prompt was provided, participants were free to choose their own topic and use the AI as much or as little as they preferred. 

Following their usage of the system, participants were asked about their overall impressions, perceived changes in their writing process, and potential integration of such tools into their regular practices. These questions aimed to capture immediate reactions and broad perceptions of each condition. They were then asked about how the interface supported or interfered with their sense of control, influenced the direction of their writing, affected their perception of authorship, and whether the final product represented their unique style and voice (RQ1). Follow-up questions used speculative comparative scenarios to understand the influence of the presented interface metaphor on their perception of agency and control (RQ2). Proposed changes included generating different lengths of output, providing more or fewer explicit controls like dropdown menus or sliders, or writing more seamlessly in the writer's style or voice. Participants were asked if these changes affected whether or not they felt like they collaborated with the AI and how in control they felt during their interaction with the AI. The full study protocol is in (\ref{sec:interview_qs}).

\subsection{Measures and Analysis}
We began with an initial open coding of the study transcripts provided by Otter to identify recurring concepts in participants' responses. This process involved listening to each interview and assigning codes to relevant segments of text, allowing us to capture the essence of participants' experiences and perspectives. Following this initial coding, we conducted a reflexive thematic analysis, as described by Braun and Clarke \cite{clarke2017thematic}, to develop and refine themes from the coded data. Coding occurred over the course of two weeks and was performed by two members of the research team. Both coders coded the data individually, developed preliminary themes, and then reviewed the codes together to develop the themes further and ensure they were empirically justified. 

Themes from professional and non-professional writers were first analyzed separately, but when few differences in the responses between the two groups were found, the two groups were merged. The themes were then reviewed and refined again through discussions between the two coders to arrive at the findings presented herein. After identifying recurring conceptions of agency and ownership (RQ1), we constructed a taxonomy based on these recurring conceptions. 

\section{Study Results}
\label{sec:results}
The findings reflected the nuance and complexity of participants' conceptions of agency and ownership when interacting with AI co-writing systems, and how those conceptions changed based on the interface metaphor presented. Below, we first describe the points of control and types of agency and ownership that were described by participants (RQ1), and later discuss how agent- and tool-like metaphors altered writers' perceptions of these systems (RQ2). 

\subsection{Conceptions of Agency and Ownership}
In this section, we describe the nuanced conceptions of agency and ownership that arose during our study. These conceptions address RQ1. 

\subsubsection{Conceptions of Control}
\label{sec:theme1}
Points of control are those aspects of the writing process that a user believes belong under their "sphere of control" \cite{Spittal2002}. Participants believe they should have a strong causal effect on the generated writing based on the particular point of control which they favored. This is relevant to \textbf{RQ2} because depending on the point of control evoked by a particular interface metaphor, the type of agency which a user feels may vary. We identified three types of agency expressed by our participants: \textit{explicit control, implicit control,} and \textit{writing process control}.

\paragraph{Explicit Control}
Some participants (i.e., P2, P6 tool-like system; P7, P10 magic-like system) felt that their sense of control lay in their \textit{explicit control} over their interaction with the AI. This was via interface elements outside the main writing area, such as buttons or dropdown menus. For these participants, being able to choose when and how they asked the AI to generate text afforded them more control: \textit{"I feel better in general if I have more control over my work. So when I change the parameter that means if I like it, I can make it better by controlling that parameter. If I don't like it then I can remove it or I can adjust the parameters. I think the flexibility leads the user to feel more confident to use the tool."} (P6, tool-like system).


\paragraph{Implicit Control}
Other participants (e.g., P1, P6, P17 tool-like system; P15, P18 agent-like system) used a form of \textit{implicit control} during their interaction: they "prompted" the AI with their previously written text. As P1 noted, \textit{"I kept trying to get it to generate a different tone, right, and to go down a different path and I felt like it took a lot of prompts."} Given that all 18 participants had the option of using the previously written text to guide the AI (regardless of the interface metaphor they were presented with), it's surprising only 5 mentioned this implicit guidance. Further, the use of the word "prompt" suggests a familiarity with existing agent-like metaphors for thinking about AI, even though three of the five participants that mentioned prompting used the tool-like system. This suggests that preconceived metaphors of AI from past experiences may have influenced how participants perceived  these systems. 

\paragraph{Writing Process Control}
For others, their ability to continue using their traditional writing process resulted in a sense of \textit{writing process control} over their interaction. These participants were able to find an internal point of control in their writing process rather than one within their interaction with the AI. For instance, participants whose writing process involved simultaneous editing and drafting did not feel that their sense of control was threatened by the AI because their usual writing process was not altered e.g., \textit{"I was obviously still in control of the piece, because I actually deleted a few sentences ...  I kept the parts I liked, I deleted the parts I didn't like, and I could still edit them."} (P11, agent-like system). P7 (magic-like system) described editing as an inevitable part of working with the AI generated text, and conceptualized their role as that of a senior partner or editor. For these writers, maintaining control in the writing process was easy regardless of the interface metaphor presented because they could still use their traditional writing process.

Although control via editing was the most frequently referenced form of writing process control, other sources of alignment and misalignment between the writer's normal process and AI co-writing process were also observed. P1 (tool-like system) described their process as an \textit{"...a sort of invitation to my subconscious to surprise me. To try and access some stuff that my conscious mind isn't able to access well"}, and therefore felt that large chunks of output from the AI violated this ability to surprise themselves. P2 (tool-like system) described their process as \textit{"aleatory and improvisational"}, and described struggling to stay in that familiar process during their interaction: \textit{"Every time it generated something it was a bit like 'oh, I wouldn't write that, I wouldn't go that way' but I still want to keep going. I want to see if it could give me something that I wanted, but obviously without me telegraphing my intentions outright"}. For writers with processes relying on randomness or reduced active intent, they may cede some control to their process. They require an AI collaborator that is receptive to their goals and style without forcing them to control it in ways that require clear intent and thus violate their writing process.


\subsubsection{Conceptions of Ownership}
\label{sec:theme2}
As with agency, participants reported a variety of ways of conceptualizing ownership over the writing. When trying to conceptualize the degree of ownership P1 (tool-like system) had over the text he generated with the AI, he stated: \textit{"It feels sort of like in the 1950s with [store bought] pancake batter when all you had to do was add water no one wanted to use it, because it felt like they were cheating so what they did is, even though you didn't need to do this, is re-advertised that you had to add an egg to it. You didn't actually have to add the egg, but the act of adding the egg was just enough for people to feel like it and had their own homemade touch."} In this section we explore how and when different participants felt they had added their "egg" to the AI generated text.

\paragraph{Stylistic Ownership}
One notable difference between the professional creative writer population and the non-professional writer population was that each tended to derive ownership from different parts of the process. Generally, non-professional writers tended to believe that conceptual contributions or "ideas" were the more important contribution and provided a sense of \textit{conceptual ownership}. Non-professional writers (e.g., P7, P16 tool-like system; P10 magic-like system, P11, P13 agent-like system) expressed this sentiment, with P11 stating \textit{ "When you're writing a book, you can have an editor, but still, the ownership and the authorship belongs to a person who's writing a book, because they contribute the ideas and they are the creative brain behind the book."} As long as the AI co-writing system was expanding on the ideas that they provided, these non-professional writers felt that they had much stronger ownership over the draft. 

Professional creative writers, in contrast, tended to place more emphasis on style, language, and the words themselves. Professional creative writer P1 (tool-like system) noted that, "\textit{I think there's a difference between feeling ownership over syntax and style versus feeling ownership over content, right?}  P1 felt that the generated text was so disparate from their own style, they could not claim any ownership over the final product. P18 (agent-like system), upon using the AI co-writing system, no longer considered themselves the author of the piece because the AI was inserting language directly into the text. P18 described the final piece as "collaborative" and compared the process to the exquisite corpse collaborative writing game\footnote{https://en.wikipedia.org/wiki/exquisite\_corpse}: \textit{"It didn't feel like a solo thing. It didn't feel like I was the writer at all times as I wasn't. Like that little game that you play when you're kids, where you say one sentence and then you have to repeat the sentence, and then add on to the sentence, and then you keep going with the story down the line."} When asked how to assess whether they or the AI had greater ownership over the text, some professional creative writers again emphasized language over ideas by suggesting that comparing the number of words that the AI composed to the number of words that the writer composed would be a reasonable way to assess ownership. For example, P8 (magic-like system) stated, \textit{"Maybe it's something to do with the amount of words taken without modification. If I had grabbed this much text and kept that I might feel like I had less of a hand in it."}  

\paragraph{Effort-Based Ownership}
Many participants (P1, P2, P14 tool-like system; P12 magic-like system) measured their ownership by the amount of energy they expended over a draft, a form of \textit{effort-based ownership}, arguing that they would not take ownership over a draft until they had spent a great deal of time editing. For example, P14 noted, \textit{"It's that it's doing the cognitive work, it would feel like it has sort of done my thinking for me. Maybe this is a thing I've internalized --- Calvinism or something. It's like you have to suffer... I have to say that I did this, that I sweated it out."} For this participant, agency and ownership were interrelated because if they were able to expend energy on points of control they considered important, then they were more likely to claim ownership over the text. 

\subsection{Interface Metaphors' Effect on Agency and Ownership}
Below, we describe the effects of the three interface metaphor systems on conceptions of agency and ownership (RQ2). 

\subsubsection{Participants Found Points of Control in Tool-Like Interface Elements} 
\label{sec:theme3}
Tool-like condition participants P2, P6 and P16 stated that they needed to have explicit control over the AI and stylistic ownership over its output to feel that the writing process was cooperative. As P16 stated, \textit{"If the AI writes seamlessly [in my style] then I'll feel very good about it, and a collaboration is going to be equal."} Likewise, P2 said that while they would tend to not use an AI co-writing system, \textit{"the more control that I can have with regard to a program like this, it lowers my resistance, shall I say, to using it."}  

Participants who used the tool-like system viewed the dropdowns and sliders that "controlled" the style, verbosity, or creativity of the generated text as important points of control (P1, P2, P6, P14, and P16). P6 stated, \textit{"I do feel better in general because I have more control over my work so when I change the parameter, that means, if I like it, I can make it better. I think the flexibility leads the user to feel more confident to use the tool.}" P2 noted \textit{"I think I would go down the rabbit hole and just play with every parameter and that's what it would take for me to kind of accept the program. I like fine grain control, wanting to see control or accept control."} For some participants, this illusion of explicit control made them feel greater agency in their interaction with the AI, while others who noticed that parameters weren't working appeared to feel less (e.g., \textit{"I don't know these sliders didn't really seem to make much difference. Halfway through I sort of tried to dial back the verbosity and creativity and it didn't really seem to do much of anything." (P17)}. 


Although the tool-like system did not alter the generated text when the control values changed, the ambiguity of the parameter terms (e.g., "Creativity" and "Style") led some participants to perceive themselves as having more control over their process. These participants reacted confidently and positively to the experimental system. For example, P6 stated, \textit{"the more control the user has, then the more confident the user feels about the text."}). Others perceived that the generated output did not match their desired style and adopted an experimental or exploratory approach where they became interested in seeing\textit{ "what the AI was going to do"} (P17). P1 drew a comparison between using the AI and approaching a synthesizer for the first time, aiming to explore what control he had over the AI rather than trying to write a complete piece: \textit{"Encountering it as a tool, my first thing with any tool---like if this was a music program---just fuck around with it."} Both P1 and P4 described ceding control to the AI during their interaction to understand its limits, aiming instead to experiment with the system.

\subsubsection{Ownership in Conceptual Contributions When Using Agent-like Interfaces}
\label{sec:theme4}

Participants that used the agent-like system tended to focus on the importance of ideation and conceptual contributions like plot points, character sketches, and setting as opposed to individual sentences. Some participants, like P13, argued that the agent-like system expanded on what they had previously written, stating, \textit{"I think [the generated text] is based on what I've already written down, and it doesn't spark many new ideas for me to write about."} P16 echoed this sentiment, stating that the AI takes what is written and \textit{"broadens the idea. It gives it expansion."} P11 also described how a stronger collaborator in a project is the one who contributes more "ideas" to the piece, \textit{"I do give Wordsworth some idea of what I want, but some ideas definitely come from the AI. That's why it's a cooperation, for sure."}

We hypothesize that participants who used the agent-like system tended to attribute ownership to conceptual contributions because the presence of an agent-like AI metaphor created the expectation that participants had the ability to converse with the AI about the piece and to communicate their desired intentions. After using the agent-like system, P9 emphasized this crucial change in dynamic, stating that \textit{"[When I work with a human collaborator] I'll call him out for stuff, and he'll be like, actually, I think that works here because of this reason and I'll be like, oh, you know, that's a great point [but an AI] can't defend those choices, and because of that, I would want it to mirror what I do."} P9's desire for the AI to defend its choices and discuss the piece with them underscores how participants who used the agent-like system valued conceptual communication when interacting with the agentic AI.  

However, because the agent-like system was only capable of fulfilling requests to generate more text, it was not the functionality of the system that spurred participants to think about conceptual contributions, but rather the interface metaphor itself. It is notable that this focus on conceptual contributions was not restricted to the desire to communicate concepts back and forth in the AI chat window, but that the agentic metaphor encouraged participants to value \textit{conceptual ownership} more broadly. This suggests that the agentic metaphor could be used to emphasize conceptual thinking in AI co-writing systems, regardless of the capabilities of the system.

\section{Taxonomy of Creative Writers' Agency and Ownership}

Our taxonomy introduces a nuanced framework for understanding creative writers' perceptions of agency and ownership when working with AI co-writing systems (\ref{tab:writer_taxonomy}). The value of this taxonomy lies in providing a unified descriptive classification of the types of agency and ownership within this domain. 

We observed three distinct types of agency that writers leverage when working with AI systems, building on and extending existing HCI frameworks. \textit{Explicit controls} \cite{Lee24}, which were instantiated differently across various interface metaphors, offer the most potential for visible, modifiable, and fine-grained control mechanisms. This echoes foundational HCI work on user control over interactive systems \cite{shneiderman81direct}. A sense of \textit{implicit control} \cite{Lee24}, exercised through the influence of the writer's text on the AI model's output, was described by many participants in our study. The conception of control as implicit echoes previous HCI findings on collaborative writing that noted how writers implicitly provided context and direction to one another during their collaboration \cite{dourish92awareness}. When implicitly exerting control over AI generated responses with their own writing before they solicit generated text from the AI, writers may draw upon previous collaborative writing experiences. \textit{Writing process control}, which has been less explored in previous HCI research, emerged as a significant source of control based on writer's cognizance of their own writing processes. Participants frequently described how having final editorial control over the generated work was a significant point of control, echoing previous findings that writers demand "final say" over AI co-writing processes \cite{reza2025co}. However, our work extends these findings by suggesting that this desire for "final say" is part of a larger conception of writing process control, which includes control derived from all alignments between AI co-writing and writers' usual creative writing processes, such as editing as they write or composing without subconscious writing goals being interrupted. 

Writers also conceived of three types of ownership: \textit{stylistic}, \textit{conceptual}, and \textit{effort-based}. Previous HCI research has explored conceptions of ownership and authorship, and our findings extend those of Draxler \cite{Draxler24} and Reza \cite{reza2025co}, building upon their foundational distinctions while introducing new empirical insights. While Draxler's work describes the relationship between authorship and ownership and Reza's framework introduces a valuable dichotomy between writers' values concerning content and form that corresponds to our findings of conceptual and stylistic types of ownership, our observational data reveals a more nuanced phenomenon specifically related to creative writing. Rather than observing ownership conceptions tied to particular populations, we found conceptions of ownership that were significantly influenced by the interface metaphor employed and the interaction dynamic this metaphor solicited. This pattern suggests that conceptual and stylistic ownership may represent universal concerns among creative writers, with variations in ownership perceptions emerging from the affordances created by particular technological interfaces and their corresponding interface metaphors.

\begin{table}[h]
\centering
\caption{Taxonomy of Agency and Ownership Types for Writers Working with AI Co-Writing Systems}
\renewcommand{\arraystretch}{1.5}
\begin{tabular}{|>{\centering\arraybackslash}m{2cm}|>{\centering\arraybackslash}m{4cm}|m{7cm}|}
\hline
\textbf{Category} & \textbf{Type}& \multicolumn{1}{|>{\centering\arraybackslash}m{7cm}|}{\textbf{Definition}} \\
\hline
\multirow{3}{2cm}[-7ex]{\centering\textbf{Agency}} & \textbf{Explicit Control} & Perception of control exercised through deliberate interface actions such as triggering AI generation, adjusting parameters, or providing direct instructions via chat, text boxes, or dropdown menus. \\
\cline{2-3}
& \textbf{Implicit Control} & Perception of control exercised through the contextual influence of previously written text on AI output, without direct instruction or interface manipulation. \\
\cline{2-3}
& \textbf{Writing Process Control} & Perception of control maintained by integrating AI contributions into established writing and revision workflows, treating generated content as material to be shaped through editing. \\
\hline
\multirow{3}{2cm}[-6ex]{\centering\textbf{Ownership}} & \textbf{Stylistic Ownership} & Perception of ownership derived from a writer's distinctive voice, language choices, and expressive elements that characterize the text. \\
\cline{2-3}
& \textbf{Conceptual Ownership} & Perception of ownership based on a writer's contribution of foundational ideas, themes, narrative direction, and creative vision. \\
\cline{2-3}
& \textbf{Effort-based Ownership} & Perception of ownership earned through the investment of cognitive effort, time, and sustained attention in developing and refining the text. \\
\hline
\end{tabular}
\label{tab:writer_taxonomy}
\end{table}



These findings also verify that Pierce et al.'s theory of psychological ownership \cite{pierce2003state} holds for AI co-writing. Pierce's framework identifies three primary routes through which individuals develop conceptions of ownership: (1) energy and labor investment, (2) intimate lived experience with the target, and (3) perception of control over the target. Our results demonstrate strong correspondence between these theoretical constructs and empirical patterns observed in AI co-writing. The first route, energy and labor investment, directly aligns with our finding of effort-based ownership and corroborates established research on the IKEA effect \cite{norton2012ikea}, which demonstrates that labor investment fosters a sense of ownership. The second route, intimate lived experience with the target, is fostered through writers' knowledge about the creative process and how the final product reflects their particular stylistic and content contributions. The third route, perception of control over the target, most readily explains the correlation between agency and ownership documented by Draxler et al. \cite{Draxler24} that manifested itself in our study by how participants exercised explicit, implicit, or process-level agency during AI co-writing.

\section{Discussion} 

In this work, we investigated agency and ownership as multivariate concepts that may be impacted by interface metaphors. By observing writers' interactions with a system instantiating one of three interface metaphor archetypes, and in follow-up interviews, we identified three types of agency and three types of ownership writers can experience during collaborations with AI co-writing systems. We also found that interface metaphors affected participants' expected sources of agency and ownership. 

While designers of interface metaphors in AI have previously focused on how mental models may be informed by interface metaphors \cite{Desai24}, our findings suggest that altering where the user expects to feel control over their interaction informs how they feel about their interaction. Because we wanted to isolate the effects of the interface metaphor, participants in both studies did not actually have different kinds of control over the system. Our interviews pointed to the fact that writers had expectations of control based on the interface metaphor, even if those expectations were not adequately met. 

Herein, we describe how our findings can help designers appropriately alter users' expectations of agency and ownership as they design and develop AI co-writing systems and discuss the limitations of our research.

\subsection{Design Recommendations For Interface Metaphors in AI Co-Writing Systems}
Three design recommendations arose from our research.
\newline

\textbf{\textit{Design Recommendation 1: Consider How Interaction Metaphors Effect Different Points of Control.}} Our findings underscored how writers find different points of control depending on their approach to writing and the affordances of the interface metaphor within an AI co-writing system. Some writers edited as they went, which led to strong editorial control over the contributions from the AI co-writing system due to their \textit{writing process control}. Others exerted \textit{explicit control} each time they invoked an AI co-writing system or in the parameters or queries used to guide the AI. A final group found \textit{implicit control} in the text they wrote and how the AI responded to their text. 

Points of control are of particular importance because they shape a writer's expectations of what a particular writing system can and should be capable of accomplishing. Our findings in [Section \ref{sec:theme3}] demonstrated that when a writer established particular points of control and those points did not meet their expectations, they were more likely to treat their interaction with an AI co-writing system as an experimental exercise rather than trying to use it to create an accomplished piece of writing. If a particular point of control is expected based on a metaphorical presentation, designers aiming for long-term adoption should ensure that the underlying system responds to a user's expectations.

Designers should also consider gathering additional information about users' writing processes before presenting them with a particular metaphor. To execute this, designers could try to solicit different points of control or conceptions of ownership depending on the metaphor presented. If a designer wants users to find their point of control in "prompting" the AI, they should present an agentic AI that emphasizes that it will be reading the previous text and providing a brief summary of what has been read before providing generated text. This change in the AI interface would emphasize to the user that the previously written text is an important input to the AI. 
\newline

\textbf{\textit{Design Recommendation 2: Design for Conceptual Relationships with Agentic AI Assistants.}} Participants that used the agent-like interface metaphor emphasized the importance of conceptual contributions more than either of the other participants. This may have been due to agents not only being conceptualized as capable of providing suggestions or editing the text, but also receiving higher-order suggestions and conversing about the text conceptually with human-like nuance \cite{Troshani21}. Designers of AI co-writing systems should decide to consciously invoke or not invoke this emphasis on higher-order thinking depending on the context. Designers seeking a collaborative dynamic, in which the user and AI collaborate by each inserting language into the final text without discussion, may opt to avoid agentic metaphors. Suppressing, modifying, or avoiding these metaphors would pave the way for users to consider the language within each sentence they compose rather than thinking of user-written sentences as conceptual fodder for the AI's responses. 
\newline

\textbf{\textit{Design Recommendation 3: Play Into or Against Expectations of Control When Using Tool-Like Metaphors.}} Our study found that tool-like metaphors set an expectation of explicit control for participants. These metaphors encouraged participants to look to other parameters, dropdown menus, and sliders as means of controlling the generated text. Therefore, it may make sense for designers to ensure that user expectations of control are met if interface parameters are presented. However, even if users cannot identify the relationship between interface controls and the generated text, this does not necessarily negatively affect their interaction. If users have difficulty mapping the ways they can alter text with the output they receive, users may treat the platform experimentally or try to generate unusual text to use in other creative output. Such text would need to be inspiring but need not be text that the writer would use instead of their own prose.

\subsection{Limitations and Future Work} 
This research has several limitations. First, the AI systems that were used in this research were informed by a review of commercial systems available today. We chose to focus on existing commercial systems rather than research prototypes to ensure our metaphors were representative of systems with which writers may already be familiar. As AI co-writing systems are rapidly evolving, new systems may be released within the next few years which introduce new metaphors. 

Also, recent work suggests that longitudinal studies of a week or more may provide more reliable results for AI creative co-writing studies \cite{carrera2025nabokov}. Further work is thus needed to explore how these results hold over longer periods. Conceptions of agency and ownership may also be influenced by particular cultural contexts, in this case North American, English-language creative writers. Future work is thus needed to explore these variables and determine the degree to which these results hold in cultures with different conceptions of ownership like East Asia, in other languages, or with other genres of writing. 


\section{Conclusion}
This paper makes several contributions to our understanding of  agency and ownership in AI co-writing systems. A review of commercial AI co-writing systems available today led to the identification of three interface metaphor archetypes (i.e., tool-like, agent-like, magic-like). Through an interview study that utilized prototypes of these archetypes, we uncovered nuanced interactions between different interface metaphors and perceptions of agency and ownership. Three types of agency and three types of ownership were identified, along with an understanding of the impact of tool-like metaphor interface controls and agent-like chat interfaces on participants' perceptions and behaviors.

Our research also provides recommendations for designers seeking to leverage interface metaphors to shape users' expectations and interactions, including considerations for different points of control, conceptual relationships with agentic AI assistants, and strategically playing into or against expectations of control with tool-like metaphors. By deepening our understanding of writers' perceptions of agency and ownership, as well as how interface metaphors mediate human-AI collaboration in writing, this work lays the groundwork for designing more effective, satisfying, and considered AI co-writing systems that meaningfully balance user agency and ownership.

\bibliographystyle{ACM-Reference-Format}
\bibliography{sample-base}


\begin{thebibliography}{66}


\ifx \showCODEN    \undefined \def \showCODEN     #1{\unskip}     \fi
\ifx \showDOI      \undefined \def \showDOI       #1{#1}\fi
\ifx \showISBNx    \undefined \def \showISBNx     #1{\unskip}     \fi
\ifx \showISBNxiii \undefined \def \showISBNxiii  #1{\unskip}     \fi
\ifx \showISSN     \undefined \def \showISSN      #1{\unskip}     \fi
\ifx \showLCCN     \undefined \def \showLCCN      #1{\unskip}     \fi
\ifx \shownote     \undefined \def \shownote      #1{#1}          \fi
\ifx \showarticletitle \undefined \def \showarticletitle #1{#1}   \fi
\ifx \showURL      \undefined \def \showURL       {\relax}        \fi
\providecommand\bibfield[2]{#2}
\providecommand\bibinfo[2]{#2}
\providecommand\natexlab[1]{#1}
\providecommand\showeprint[2][]{arXiv:#2}

\bibitem[Agre(1997)]%
        {agre1997computation}
\bibfield{author}{\bibinfo{person}{Philip Agre}.} \bibinfo{year}{1997}\natexlab{}.
\newblock \bibinfo{booktitle}{\emph{Computation and human experience}}.
\newblock \bibinfo{publisher}{Cambridge University Press}.
\newblock


\bibitem[Asmelash(2023)]%
        {cnnTheseBooks}
\bibfield{author}{\bibinfo{person}{Leah Asmelash}.} \bibinfo{year}{2023}\natexlab{}.
\newblock \bibinfo{title}{These books are being used to train AI. No one told the authors}.
\newblock \bibinfo{howpublished}{\url{https://www.cnn.com/2023/10/08/style/ai-books3-authors-nora-roberts-cec/index.html}}.
\newblock
\newblock
\shownote{[Accessed 03-09-2024]}.


\bibitem[Barthes(2016)]%
        {barthes2016death}
\bibfield{author}{\bibinfo{person}{Roland Barthes}.} \bibinfo{year}{2016}\natexlab{}.
\newblock \showarticletitle{The death of the author}.
\newblock In \bibinfo{booktitle}{\emph{Readings in the Theory of Religion}}. \bibinfo{publisher}{Routledge}, \bibinfo{pages}{141--145}.
\newblock


\bibitem[Benharrak et~al\mbox{.}(2024)]%
        {benharrakWriterDefinedAIPersonas2024}
\bibfield{author}{\bibinfo{person}{Karim Benharrak}, \bibinfo{person}{Tim Zindulka}, \bibinfo{person}{Florian Lehmann}, \bibinfo{person}{Hendrik Heuer}, {and} \bibinfo{person}{Daniel Buschek}.} \bibinfo{year}{2024}\natexlab{}.
\newblock \showarticletitle{Writer-{{Defined AI Personas}} for {{On-Demand Feedback Generation}}}. In \bibinfo{booktitle}{\emph{Proceedings of the {{CHI Conference}} on {{Human Factors}} in {{Computing Systems}}}}. \bibinfo{pages}{1--18}.
\newblock
\urldef\tempurl%
\url{https://doi.org/10.1145/3613904.3642406}
\showDOI{\tempurl}
\showeprint[arxiv]{2309.10433}~[cs]


\bibitem[Bentvelzen et~al\mbox{.}(2022)]%
        {Bentvelzen22}
\bibfield{author}{\bibinfo{person}{Marit Bentvelzen}, \bibinfo{person}{Pawe\l{}~W. Wo\'{z}niak}, \bibinfo{person}{Pia~S.F. Herbes}, \bibinfo{person}{Evropi Stefanidi}, {and} \bibinfo{person}{Jasmin Niess}.} \bibinfo{year}{2022}\natexlab{}.
\newblock \showarticletitle{Revisiting Reflection in HCI: Four Design Resources for Technologies that Support Reflection}.
\newblock \bibinfo{journal}{\emph{Proc. ACM Interact. Mob. Wearable Ubiquitous Technol.}} \bibinfo{volume}{6}, \bibinfo{number}{1}, Article \bibinfo{articleno}{2} (\bibinfo{date}{mar} \bibinfo{year}{2022}), \bibinfo{numpages}{27}~pages.
\newblock
\urldef\tempurl%
\url{https://doi.org/10.1145/3517233}
\showDOI{\tempurl}


\bibitem[Biermann et~al\mbox{.}(2022)]%
        {Biermann22}
\bibfield{author}{\bibinfo{person}{Oloff~C. Biermann}, \bibinfo{person}{Ning~F. Ma}, {and} \bibinfo{person}{Dongwook Yoon}.} \bibinfo{year}{2022}\natexlab{}.
\newblock \showarticletitle{From Tool to Companion: Storywriters Want AI Writers to Respect Their Personal Values and Writing Strategies}. In \bibinfo{booktitle}{\emph{Proceedings of the 2022 ACM Designing Interactive Systems Conference}} (, Virtual Event, Australia,) \emph{(\bibinfo{series}{DIS '22})}. \bibinfo{publisher}{Association for Computing Machinery}, \bibinfo{address}{New York, NY, USA}, \bibinfo{pages}{1209–1227}.
\newblock
\showISBNx{9781450393584}
\urldef\tempurl%
\url{https://doi.org/10.1145/3532106.3533506}
\showDOI{\tempurl}


\bibitem[Calderwood et~al\mbox{.}(2018)]%
        {calderwoodHowNovelistsUse2018}
\bibfield{author}{\bibinfo{person}{Alex Calderwood}, \bibinfo{person}{Vivian Qiu}, \bibinfo{person}{Katy~Ilonka Gero}, {and} \bibinfo{person}{Lydia~B Chilton}.} \bibinfo{year}{2018}\natexlab{}.
\newblock \showarticletitle{How {{Novelists Use Generative Language Models}}: {{An Exploratory User Study}}}. In \bibinfo{booktitle}{\emph{23rd {{International Conference}} on {{Intelligent User Interfaces}}}}. \bibinfo{publisher}{ACM}.
\newblock
\showISBNx{978-1-4503-4945-1}


\bibitem[Carrera et~al\mbox{.}(2025)]%
        {carrera2025nabokov}
\bibfield{author}{\bibinfo{person}{Dashiel Carrera}, \bibinfo{person}{Zixin Zhao}, \bibinfo{person}{Ashish Ajin~Thomas}, {and} \bibinfo{person}{Daniel Wigdor}.} \bibinfo{year}{2025}\natexlab{}.
\newblock \showarticletitle{Nabokov's Cards: An AI Assisted Prewriting System to Support Bottom-Up Creative Writing}. In \bibinfo{booktitle}{\emph{Proceedings of the 2025 Conference on Creativity and Cognition}}. \bibinfo{pages}{546--559}.
\newblock


\bibitem[Carroll et~al\mbox{.}(1988)]%
        {Carroll1988Interface}
\bibfield{author}{\bibinfo{person}{John~M. Carroll}, \bibinfo{person}{Robert~L. Mack}, {and} \bibinfo{person}{Wayne~A. Kellogg}.} \bibinfo{year}{1988}\natexlab{}.
\newblock \showarticletitle{Interface Metaphors and User Interface Design}.
\newblock In \bibinfo{booktitle}{\emph{Handbook of Human-Computer Interaction}}, \bibfield{editor}{\bibinfo{person}{M.~Helander}} (Ed.). \bibinfo{publisher}{Elsevier Science}, Chapter~8, \bibinfo{pages}{67--85}.
\newblock
\urldef\tempurl%
\url{https://doi.org/10.1016/B978-0-444-70536-5.50008-7}
\showDOI{\tempurl}


\bibitem[Chakrabarty et~al\mbox{.}(2024)]%
        {chakrabarty2024artoratifice}
\bibfield{author}{\bibinfo{person}{Tuhin Chakrabarty}, \bibinfo{person}{Philippe Laban}, \bibinfo{person}{Divyansh Agarwal}, \bibinfo{person}{Smaranda Muresan}, {and} \bibinfo{person}{Chien-Sheng Wu}.} \bibinfo{year}{2024}\natexlab{}.
\newblock \showarticletitle{Art or Artifice? Large Language Models and the False Promise of Creativity}. In \bibinfo{booktitle}{\emph{Proceedings of the 2024 CHI Conference on Human Factors in Computing Systems}} (Honolulu, HI, USA) \emph{(\bibinfo{series}{CHI '24})}. \bibinfo{publisher}{Association for Computing Machinery}, \bibinfo{address}{New York, NY, USA}, Article \bibinfo{articleno}{30}, \bibinfo{numpages}{34}~pages.
\newblock
\showISBNx{9798400703300}
\urldef\tempurl%
\url{https://doi.org/10.1145/3613904.3642731}
\showDOI{\tempurl}


\bibitem[Chin et~al\mbox{.}(2024)]%
        {ChinDesai24}
\bibfield{author}{\bibinfo{person}{Jessie Chin}, \bibinfo{person}{Smit Desai}, \bibinfo{person}{Sheny (Cheng-Hsuan) Lin}, {and} \bibinfo{person}{Shannon Mejia}.} \bibinfo{year}{2024}\natexlab{}.
\newblock \showarticletitle{Like My Aunt Dorothy: Effects of Conversational Styles on Perceptions, Acceptance and Metaphorical Descriptions of Voice Assistants during Later Adulthood}.
\newblock \bibinfo{journal}{\emph{Proc. ACM Hum.-Comput. Interact.}} \bibinfo{volume}{8}, \bibinfo{number}{CSCW1}, Article \bibinfo{articleno}{88} (\bibinfo{date}{apr} \bibinfo{year}{2024}), \bibinfo{numpages}{21}~pages.
\newblock
\urldef\tempurl%
\url{https://doi.org/10.1145/3637365}
\showDOI{\tempurl}


\bibitem[Chung and Kreminski(2024)]%
        {chung2024PatchviewLLMPoweredWorldbuilding}
\bibfield{author}{\bibinfo{person}{John Joon~Young Chung} {and} \bibinfo{person}{Max Kreminski}.} \bibinfo{year}{2024}\natexlab{}.
\newblock \bibinfo{title}{Patchview: {{LLM-Powered Worldbuilding}} with {{Generative Dust}} and {{Magnet Visualization}}}.
\newblock
\newblock
\urldef\tempurl%
\url{https://doi.org/10.1145/3654777.3676352}
\showDOI{\tempurl}
\showeprint[arxiv]{2408.04112}~[cs]


\bibitem[Ciger et~al\mbox{.}(2003)]%
        {Ciger03}
\bibfield{author}{\bibinfo{person}{Jan Ciger}, \bibinfo{person}{Mario Gutierrez}, \bibinfo{person}{Frederic Vexo}, {and} \bibinfo{person}{Daniel Thalmann}.} \bibinfo{year}{2003}\natexlab{}.
\newblock \showarticletitle{The magic wand}. In \bibinfo{booktitle}{\emph{Proceedings of the 19th Spring Conference on Computer Graphics}} (Budmerice, Slovakia) \emph{(\bibinfo{series}{SCCG '03})}. \bibinfo{publisher}{Association for Computing Machinery}, \bibinfo{address}{New York, NY, USA}, \bibinfo{pages}{119–124}.
\newblock
\showISBNx{158113861X}
\urldef\tempurl%
\url{https://doi.org/10.1145/984952.984972}
\showDOI{\tempurl}


\bibitem[Clarke and Braun(2017)]%
        {clarke2017thematic}
\bibfield{author}{\bibinfo{person}{Victoria Clarke} {and} \bibinfo{person}{Virginia Braun}.} \bibinfo{year}{2017}\natexlab{}.
\newblock \showarticletitle{Thematic analysis}.
\newblock \bibinfo{journal}{\emph{The journal of positive psychology}} \bibinfo{volume}{12}, \bibinfo{number}{3} (\bibinfo{year}{2017}), \bibinfo{pages}{297--298}.
\newblock


\bibitem[Desai et~al\mbox{.}(2024)]%
        {Desai24}
\bibfield{author}{\bibinfo{person}{Smit Desai}, \bibinfo{person}{Mateusz Dubiel}, {and} \bibinfo{person}{Luis~A. Leiva}.} \bibinfo{year}{2024}\natexlab{}.
\newblock \showarticletitle{Examining Humanness as a Metaphor to Design Voice User Interfaces}. In \bibinfo{booktitle}{\emph{Proceedings of the 6th ACM Conference on Conversational User Interfaces}} (Luxembourg, Luxembourg) \emph{(\bibinfo{series}{CUI '24})}. \bibinfo{publisher}{Association for Computing Machinery}, \bibinfo{address}{New York, NY, USA}, Article \bibinfo{articleno}{7}, \bibinfo{numpages}{15}~pages.
\newblock
\showISBNx{9798400705113}
\urldef\tempurl%
\url{https://doi.org/10.1145/3640794.3665535}
\showDOI{\tempurl}


\bibitem[Desai and Twidale(2023)]%
        {Desai23}
\bibfield{author}{\bibinfo{person}{Smit Desai} {and} \bibinfo{person}{Michael Twidale}.} \bibinfo{year}{2023}\natexlab{}.
\newblock \showarticletitle{Metaphors in Voice User Interfaces: A Slippery Fish}.
\newblock \bibinfo{journal}{\emph{ACM Trans. Comput.-Hum. Interact.}} \bibinfo{volume}{30}, \bibinfo{number}{6}, Article \bibinfo{articleno}{89} (\bibinfo{date}{sep} \bibinfo{year}{2023}), \bibinfo{numpages}{37}~pages.
\newblock
\showISSN{1073-0516}
\urldef\tempurl%
\url{https://doi.org/10.1145/3609326}
\showDOI{\tempurl}


\bibitem[Dhillon et~al\mbox{.}(2024)]%
        {dhillon2024ShapingHumanAICollaboration}
\bibfield{author}{\bibinfo{person}{Paramveer~S. Dhillon}, \bibinfo{person}{Somayeh Molaei}, \bibinfo{person}{Jiaqi Li}, \bibinfo{person}{Maximilian Golub}, \bibinfo{person}{Shaochun Zheng}, {and} \bibinfo{person}{Lionel~Peter Robert}.} \bibinfo{year}{2024}\natexlab{}.
\newblock \showarticletitle{Shaping {{Human-AI Collaboration}}: {{Varied Scaffolding Levels}} in {{Co-writing}} with {{Language Models}}}. In \bibinfo{booktitle}{\emph{Proceedings of the {{CHI Conference}} on {{Human Factors}} in {{Computing Systems}}}}. \bibinfo{publisher}{ACM}, \bibinfo{address}{Honolulu HI USA}, \bibinfo{pages}{1--18}.
\newblock
\showISBNx{9798400703300}
\urldef\tempurl%
\url{https://doi.org/10.1145/3613904.3642134}
\showDOI{\tempurl}


\bibitem[Dourish and Bellotti(1992)]%
        {dourish92awareness}
\bibfield{author}{\bibinfo{person}{Paul Dourish} {and} \bibinfo{person}{Victoria Bellotti}.} \bibinfo{year}{1992}\natexlab{}.
\newblock \showarticletitle{Awareness and coordination in shared workspaces}. In \bibinfo{booktitle}{\emph{Proceedings of the 1992 ACM Conference on Computer-Supported Cooperative Work}} (Toronto, Ontario, Canada) \emph{(\bibinfo{series}{CSCW '92})}. \bibinfo{publisher}{Association for Computing Machinery}, \bibinfo{address}{New York, NY, USA}, \bibinfo{pages}{107–114}.
\newblock
\showISBNx{0897915429}
\urldef\tempurl%
\url{https://doi.org/10.1145/143457.143468}
\showDOI{\tempurl}


\bibitem[Draxler et~al\mbox{.}(2024)]%
        {Draxler24}
\bibfield{author}{\bibinfo{person}{Fiona Draxler}, \bibinfo{person}{Anna Werner}, \bibinfo{person}{Florian Lehmann}, \bibinfo{person}{Matthias Hoppe}, \bibinfo{person}{Albrecht Schmidt}, \bibinfo{person}{Daniel Buschek}, {and} \bibinfo{person}{Robin Welsch}.} \bibinfo{year}{2024}\natexlab{}.
\newblock \showarticletitle{The AI Ghostwriter Effect: When Users do not Perceive Ownership of AI-Generated Text but Self-Declare as Authors}.
\newblock \bibinfo{journal}{\emph{ACM Trans. Comput.-Hum. Interact.}} \bibinfo{volume}{31}, \bibinfo{number}{2}, Article \bibinfo{articleno}{25} (\bibinfo{date}{feb} \bibinfo{year}{2024}), \bibinfo{numpages}{40}~pages.
\newblock
\showISSN{1073-0516}
\urldef\tempurl%
\url{https://doi.org/10.1145/3637875}
\showDOI{\tempurl}


\bibitem[DreamGen(2024)]%
        {dreamgen}
\bibfield{author}{\bibinfo{person}{DreamGen}.} \bibinfo{year}{2024}\natexlab{}.
\newblock \bibinfo{title}{AI Role-Play and Story Generator | DreamGen}.
\newblock \bibinfo{howpublished}{\url{https://dreamgen.com/}}.
\newblock
\newblock
\shownote{Retrieved: 2024-06-03}.


\bibitem[Gero and Chilton(2019)]%
        {geroMetaphoriaAlgorithmicCompanion2019}
\bibfield{author}{\bibinfo{person}{Katy~Ilonka Gero} {and} \bibinfo{person}{Lydia~B. Chilton}.} \bibinfo{year}{2019}\natexlab{}.
\newblock \showarticletitle{Metaphoria: {{An Algorithmic Companion}} for {{Metaphor Creation}}}. In \bibinfo{booktitle}{\emph{Proceedings of the 2019 {{CHI Conference}} on {{Human Factors}} in {{Computing Systems}}}}. \bibinfo{publisher}{ACM}, \bibinfo{address}{Glasgow Scotland Uk}, \bibinfo{pages}{1--12}.
\newblock
\showISBNx{978-1-4503-5970-2}
\urldef\tempurl%
\url{https://doi.org/10.1145/3290605.3300526}
\showDOI{\tempurl}


\bibitem[Gero et~al\mbox{.}(2022)]%
        {geroSparksInspirationScience2022}
\bibfield{author}{\bibinfo{person}{Katy~Ilonka Gero}, \bibinfo{person}{Vivian Liu}, {and} \bibinfo{person}{Lydia Chilton}.} \bibinfo{year}{2022}\natexlab{}.
\newblock \showarticletitle{Sparks: {{Inspiration}} for {{Science Writing}} Using {{Language Models}}}. In \bibinfo{booktitle}{\emph{Designing {{Interactive Systems Conference}}}}. \bibinfo{publisher}{ACM}, \bibinfo{address}{Virtual Event Australia}, \bibinfo{pages}{1002--1019}.
\newblock
\showISBNx{978-1-4503-9358-4}
\urldef\tempurl%
\url{https://doi.org/10.1145/3532106.3533533}
\showDOI{\tempurl}


\bibitem[Gero et~al\mbox{.}(2023)]%
        {Gero23}
\bibfield{author}{\bibinfo{person}{Katy~Ilonka Gero}, \bibinfo{person}{Tao Long}, {and} \bibinfo{person}{Lydia~B Chilton}.} \bibinfo{year}{2023}\natexlab{}.
\newblock \showarticletitle{Social Dynamics of AI Support in Creative Writing}. In \bibinfo{booktitle}{\emph{Proceedings of the 2023 CHI Conference on Human Factors in Computing Systems}} (Hamburg, Germany) \emph{(\bibinfo{series}{CHI '23})}. \bibinfo{publisher}{Association for Computing Machinery}, \bibinfo{address}{New York, NY, USA}, Article \bibinfo{articleno}{245}, \bibinfo{numpages}{15}~pages.
\newblock
\showISBNx{9781450394215}
\urldef\tempurl%
\url{https://doi.org/10.1145/3544548.3580782}
\showDOI{\tempurl}


\bibitem[Ghajargar et~al\mbox{.}(2022)]%
        {Ghajargar22}
\bibfield{author}{\bibinfo{person}{Maliheh Ghajargar}, \bibinfo{person}{Jeffrey Bardzell}, {and} \bibinfo{person}{Love Lagerkvist}.} \bibinfo{year}{2022}\natexlab{}.
\newblock \showarticletitle{A Redhead Walks into a Bar: Experiences of Writing Fiction with Artificial Intelligence}. In \bibinfo{booktitle}{\emph{Proceedings of the 25th International Academic Mindtrek Conference}} (Tampere, Finland) \emph{(\bibinfo{series}{Academic Mindtrek '22})}. \bibinfo{publisher}{Association for Computing Machinery}, \bibinfo{address}{New York, NY, USA}, \bibinfo{pages}{230–241}.
\newblock
\showISBNx{9781450399555}
\urldef\tempurl%
\url{https://doi.org/10.1145/3569219.3569418}
\showDOI{\tempurl}


\bibitem[G\"{u}ldenpfennig et~al\mbox{.}(2019)]%
        {Guldenpfennig19}
\bibfield{author}{\bibinfo{person}{Florian G\"{u}ldenpfennig}, \bibinfo{person}{Daniel Dudo}, {and} \bibinfo{person}{Peter Purgathofer}.} \bibinfo{year}{2019}\natexlab{}.
\newblock \showarticletitle{The 'Magic Paradigm' for Programming Smart Connected Devices}. In \bibinfo{booktitle}{\emph{Extended Abstracts of the 2019 CHI Conference on Human Factors in Computing Systems}} (Glasgow, Scotland Uk) \emph{(\bibinfo{series}{CHI EA '19})}. \bibinfo{publisher}{Association for Computing Machinery}, \bibinfo{address}{New York, NY, USA}, \bibinfo{pages}{1–6}.
\newblock
\showISBNx{9781450359719}
\urldef\tempurl%
\url{https://doi.org/10.1145/3290607.3312892}
\showDOI{\tempurl}


\bibitem[Guo et~al\mbox{.}(2024)]%
        {guo2024ExploringImpactAI}
\bibfield{author}{\bibinfo{person}{Alicia Guo}, \bibinfo{person}{Pat Pataranutaporn}, {and} \bibinfo{person}{Pattie Maes}.} \bibinfo{year}{2024}\natexlab{}.
\newblock \showarticletitle{Exploring the {{Impact}} of {{AI Value Alignment}} in {{Collaborative Ideation}}: {{Effects}} on {{Perception}}, {{Ownership}}, and {{Output}}}. In \bibinfo{booktitle}{\emph{Extended {{Abstracts}} of the {{CHI Conference}} on {{Human Factors}} in {{Computing Systems}}}}. \bibinfo{publisher}{ACM}, \bibinfo{address}{Honolulu HI USA}, \bibinfo{pages}{1--11}.
\newblock
\showISBNx{9798400703317}
\urldef\tempurl%
\url{https://doi.org/10.1145/3613905.3650892}
\showDOI{\tempurl}


\bibitem[Hoque et~al\mbox{.}(2024)]%
        {hoque2024HaLLMarkEffectSupporting}
\bibfield{author}{\bibinfo{person}{Md~Naimul Hoque}, \bibinfo{person}{Tasfia Mashiat}, \bibinfo{person}{Bhavya Ghai}, \bibinfo{person}{Cecilia~D. Shelton}, \bibinfo{person}{Fanny Chevalier}, \bibinfo{person}{Kari Kraus}, {and} \bibinfo{person}{Niklas Elmqvist}.} \bibinfo{year}{2024}\natexlab{}.
\newblock \showarticletitle{The {{HaLLMark Effect}}: {{Supporting Provenance}} and {{Transparent Use}} of {{Large Language Models}} in {{Writing}} with {{Interactive Visualization}}}. In \bibinfo{booktitle}{\emph{Proceedings of the {{CHI Conference}} on {{Human Factors}} in {{Computing Systems}}}}. \bibinfo{publisher}{ACM}, \bibinfo{address}{Honolulu HI USA}, \bibinfo{pages}{1--15}.
\newblock
\showISBNx{9798400703300}
\urldef\tempurl%
\url{https://doi.org/10.1145/3613904.3641895}
\showDOI{\tempurl}


\bibitem[Hui and See(2015)]%
        {Hui15}
\bibfield{author}{\bibinfo{person}{Sarah Low~Tze Hui} {and} \bibinfo{person}{Swee~Lan See}.} \bibinfo{year}{2015}\natexlab{}.
\newblock \showarticletitle{Enhancing User Experience Through Customisation of UI Design}.
\newblock \bibinfo{journal}{\emph{Procedia Manufacturing}}  \bibinfo{volume}{3} (\bibinfo{year}{2015}), \bibinfo{pages}{1932--1937}.
\newblock
\showISSN{2351-9789}
\urldef\tempurl%
\url{https://doi.org/10.1016/j.promfg.2015.07.237}
\showDOI{\tempurl}
\newblock
\shownote{6th International Conference on Applied Human Factors and Ergonomics (AHFE 2015) and the Affiliated Conferences, AHFE 2015}.


\bibitem[Ippolito et~al\mbox{.}(2022)]%
        {ippolitoWordCraftCreativeWritingAIPowered2022}
\bibfield{author}{\bibinfo{person}{Daphne Ippolito}, \bibinfo{person}{Ann Yuan}, \bibinfo{person}{Andy Coenen}, {and} \bibinfo{person}{Sehmon Burnam}.} \bibinfo{year}{2022}\natexlab{}.
\newblock \bibinfo{title}{Creative {{Writing}} with an {{AI-Powered Writing Assistant}}: {{Perspectives}} from {{Professional Writers}}}.
\newblock
\newblock
\showeprint[arxiv]{2211.05030}~[cs]


\bibitem[Kadoma et~al\mbox{.}(2024)]%
        {kadoma2024RoleInclusionControl}
\bibfield{author}{\bibinfo{person}{Kowe Kadoma}, \bibinfo{person}{Marianne Aubin Le~Quere}, \bibinfo{person}{Xiyu~Jenny Fu}, \bibinfo{person}{Christin Munsch}, \bibinfo{person}{Dana{\"e} Metaxa}, {and} \bibinfo{person}{Mor Naaman}.} \bibinfo{year}{2024}\natexlab{}.
\newblock \showarticletitle{The {{Role}} of {{Inclusion}}, {{Control}}, and {{Ownership}} in {{Workplace AI-Mediated Communication}}}. In \bibinfo{booktitle}{\emph{Proceedings of the {{CHI Conference}} on {{Human Factors}} in {{Computing Systems}}}}. \bibinfo{publisher}{ACM}, \bibinfo{address}{Honolulu HI USA}, \bibinfo{pages}{1--10}.
\newblock
\showISBNx{9798400703300}
\urldef\tempurl%
\url{https://doi.org/10.1145/3613904.3642650}
\showDOI{\tempurl}


\bibitem[Kariyawasam et~al\mbox{.}(2024)]%
        {kariyawasam2024AppropriateIncongruityDriven}
\bibfield{author}{\bibinfo{person}{Hasindu Kariyawasam}, \bibinfo{person}{Amashi Niwarthana}, \bibinfo{person}{Alister Palmer}, \bibinfo{person}{Judy Kay}, {and} \bibinfo{person}{Anusha Withana}.} \bibinfo{year}{2024}\natexlab{}.
\newblock \showarticletitle{Appropriate {{Incongruity Driven Human-AI Collaborative Tool}} to {{Assist Novices}} in {{Humorous Content Generation}}}. In \bibinfo{booktitle}{\emph{Proceedings of the 29th {{International Conference}} on {{Intelligent User Interfaces}}}}. \bibinfo{publisher}{ACM}, \bibinfo{address}{Greenville SC USA}, \bibinfo{pages}{650--659}.
\newblock
\showISBNx{9798400705083}
\urldef\tempurl%
\url{https://doi.org/10.1145/3640543.3645161}
\showDOI{\tempurl}


\bibitem[Khadpe et~al\mbox{.}(2020)]%
        {Khadpe20}
\bibfield{author}{\bibinfo{person}{Pranav Khadpe}, \bibinfo{person}{Ranjay Krishna}, \bibinfo{person}{Li Fei-Fei}, \bibinfo{person}{Jeffrey~T. Hancock}, {and} \bibinfo{person}{Michael~S. Bernstein}.} \bibinfo{year}{2020}\natexlab{}.
\newblock \showarticletitle{Conceptual Metaphors Impact Perceptions of Human-AI Collaboration}.
\newblock \bibinfo{journal}{\emph{Proc. ACM Hum.-Comput. Interact.}} \bibinfo{volume}{4}, \bibinfo{number}{CSCW2}, Article \bibinfo{articleno}{163} (\bibinfo{date}{oct} \bibinfo{year}{2020}), \bibinfo{numpages}{26}~pages.
\newblock
\urldef\tempurl%
\url{https://doi.org/10.1145/3415234}
\showDOI{\tempurl}


\bibitem[Kim et~al\mbox{.}(2024)]%
        {kim2024AuthorsValuesAttitudes}
\bibfield{author}{\bibinfo{person}{Taewook Kim}, \bibinfo{person}{Hyomin Han}, \bibinfo{person}{Eytan Adar}, \bibinfo{person}{Matthew Kay}, {and} \bibinfo{person}{John Joon~Young Chung}.} \bibinfo{year}{2024}\natexlab{}.
\newblock \showarticletitle{Authors' {{Values}} and {{Attitudes Towards AI-bridged Scalable Personalization}} of {{Creative Language Arts}}}. In \bibinfo{booktitle}{\emph{Proceedings of the {{CHI Conference}} on {{Human Factors}} in {{Computing Systems}}}}. \bibinfo{publisher}{ACM}, \bibinfo{address}{Honolulu HI USA}, \bibinfo{pages}{1--16}.
\newblock
\showISBNx{9798400703300}
\urldef\tempurl%
\url{https://doi.org/10.1145/3613904.3642529}
\showDOI{\tempurl}


\bibitem[Kobiella et~al\mbox{.}(2024)]%
        {kobiella2024IfMachineGood}
\bibfield{author}{\bibinfo{person}{Charlotte Kobiella}, \bibinfo{person}{Yarhy~Said Flores~L{\'o}pez}, \bibinfo{person}{Franz Waltenberger}, \bibinfo{person}{Fiona Draxler}, {and} \bibinfo{person}{Albrecht Schmidt}.} \bibinfo{year}{2024}\natexlab{}.
\newblock \showarticletitle{"{{If}} the {{Machine Is As Good As Me}}, {{Then What Use Am I}}?" -- {{How}} the {{Use}} of {{ChatGPT Changes Young Professionals}}' {{Perception}} of {{Productivity}} and {{Accomplishment}}}. In \bibinfo{booktitle}{\emph{Proceedings of the {{CHI Conference}} on {{Human Factors}} in {{Computing Systems}}}}. \bibinfo{publisher}{ACM}, \bibinfo{address}{Honolulu HI USA}, \bibinfo{pages}{1--16}.
\newblock
\showISBNx{9798400703300}
\urldef\tempurl%
\url{https://doi.org/10.1145/3613904.3641964}
\showDOI{\tempurl}


\bibitem[Lee et~al\mbox{.}(2024)]%
        {Lee24}
\bibfield{author}{\bibinfo{person}{Mina Lee}, \bibinfo{person}{Katy~Ilonka Gero}, \bibinfo{person}{John Joon~Young Chung}, \bibinfo{person}{Simon~Buckingham Shum}, \bibinfo{person}{Vipul Raheja}, \bibinfo{person}{Hua Shen}, \bibinfo{person}{Subhashini Venugopalan}, \bibinfo{person}{Thiemo Wambsganss}, \bibinfo{person}{David Zhou}, \bibinfo{person}{Emad~A. Alghamdi}, \bibinfo{person}{Tal August}, \bibinfo{person}{Avinash Bhat}, \bibinfo{person}{Madiha~Zahrah Choksi}, \bibinfo{person}{Senjuti Dutta}, \bibinfo{person}{Jin~L.C. Guo}, \bibinfo{person}{Md~Naimul Hoque}, \bibinfo{person}{Yewon Kim}, \bibinfo{person}{Simon Knight}, \bibinfo{person}{Seyed~Parsa Neshaei}, \bibinfo{person}{Antonette Shibani}, \bibinfo{person}{Disha Shrivastava}, \bibinfo{person}{Lila Shroff}, \bibinfo{person}{Agnia Sergeyuk}, \bibinfo{person}{Jessi Stark}, \bibinfo{person}{Sarah Sterman}, \bibinfo{person}{Sitong Wang}, \bibinfo{person}{Antoine Bosselut}, \bibinfo{person}{Daniel Buschek}, \bibinfo{person}{Joseph~Chee Chang},
  \bibinfo{person}{Sherol Chen}, \bibinfo{person}{Max Kreminski}, \bibinfo{person}{Joonsuk Park}, \bibinfo{person}{Roy Pea}, \bibinfo{person}{Eugenia Ha~Rim Rho}, \bibinfo{person}{Zejiang Shen}, {and} \bibinfo{person}{Pao Siangliulue}.} \bibinfo{year}{2024}\natexlab{}.
\newblock \showarticletitle{A Design Space for Intelligent and Interactive Writing Assistants}. In \bibinfo{booktitle}{\emph{Proceedings of the CHI Conference on Human Factors in Computing Systems}} (Honolulu, HI, USA) \emph{(\bibinfo{series}{CHI '24})}. \bibinfo{publisher}{Association for Computing Machinery}, \bibinfo{address}{New York, NY, USA}, Article \bibinfo{articleno}{1054}, \bibinfo{numpages}{35}~pages.
\newblock
\showISBNx{9798400703300}
\urldef\tempurl%
\url{https://doi.org/10.1145/3613904.3642697}
\showDOI{\tempurl}


\bibitem[Lee et~al\mbox{.}(2022)]%
        {Lee22}
\bibfield{author}{\bibinfo{person}{Mina Lee}, \bibinfo{person}{Percy Liang}, {and} \bibinfo{person}{Qian Yang}.} \bibinfo{year}{2022}\natexlab{}.
\newblock \showarticletitle{CoAuthor: Designing a Human-AI Collaborative Writing Dataset for Exploring Language Model Capabilities}. In \bibinfo{booktitle}{\emph{Proceedings of the 2022 CHI Conference on Human Factors in Computing Systems}} (New Orleans, LA, USA) \emph{(\bibinfo{series}{CHI '22})}. \bibinfo{publisher}{Association for Computing Machinery}, \bibinfo{address}{New York, NY, USA}, Article \bibinfo{articleno}{388}, \bibinfo{numpages}{19}~pages.
\newblock
\showISBNx{9781450391573}
\urldef\tempurl%
\url{https://doi.org/10.1145/3491102.3502030}
\showDOI{\tempurl}


\bibitem[Li and Luximon(2022)]%
        {Li2022Navigating}
\bibfield{author}{\bibinfo{person}{Qingchuan Li} {and} \bibinfo{person}{Yan Luximon}.} \bibinfo{year}{2022}\natexlab{}.
\newblock \showarticletitle{Navigating the Mobile Applications: The Influence of Interface Metaphor and Other Factors on Older Adults’ Navigation Behavior}.
\newblock \bibinfo{journal}{\emph{International Journal of Human-Computer Interaction}} (\bibinfo{year}{2022}).
\newblock
\urldef\tempurl%
\url{https://doi.org/10.1080/10447318.2022.2050540}
\showDOI{\tempurl}


\bibitem[Lupetti and Murray-Rust(2024)]%
        {Lupetti24}
\bibfield{author}{\bibinfo{person}{Maria~Luce Lupetti} {and} \bibinfo{person}{Dave Murray-Rust}.} \bibinfo{year}{2024}\natexlab{}.
\newblock \showarticletitle{(Un)making AI Magic: A Design Taxonomy}. In \bibinfo{booktitle}{\emph{Proceedings of the CHI Conference on Human Factors in Computing Systems}} (Honolulu, HI, USA) \emph{(\bibinfo{series}{CHI '24})}. \bibinfo{publisher}{Association for Computing Machinery}, \bibinfo{address}{New York, NY, USA}, Article \bibinfo{articleno}{1}, \bibinfo{numpages}{21}~pages.
\newblock
\showISBNx{9798400703300}
\urldef\tempurl%
\url{https://doi.org/10.1145/3613904.3641954}
\showDOI{\tempurl}


\bibitem[Marche(2023)]%
        {marche2023death}
\bibfield{author}{\bibinfo{person}{Stephen Marche}.} \bibinfo{year}{2023}\natexlab{}.
\newblock \showarticletitle{Death of an Author: AI Wrote 95 Percent of This Murder Mystery}.
\newblock \bibinfo{journal}{\emph{Wired}} (\bibinfo{date}{May 15} \bibinfo{year}{2023}).
\newblock
\newblock
\shownote{Retrieved from \url{https://www.wired.com/story/death-of-an-author-ai-book-review/}}.


\bibitem[Marcus(1992)]%
        {marcus1992graphic}
\bibfield{author}{\bibinfo{person}{Aaron Marcus}.} \bibinfo{year}{1992}\natexlab{}.
\newblock \bibinfo{booktitle}{\emph{Graphic Design for Electronic Documents and User Interfaces}}.
\newblock \bibinfo{publisher}{Addison-Wesley}.
\newblock


\bibitem[Marshall et~al\mbox{.}(2010)]%
        {Marshall10}
\bibfield{author}{\bibinfo{person}{Joe Marshall}, \bibinfo{person}{Steve Benford}, {and} \bibinfo{person}{Tony Pridmore}.} \bibinfo{year}{2010}\natexlab{}.
\newblock \showarticletitle{Deception and magic in collaborative interaction}. In \bibinfo{booktitle}{\emph{Proceedings of the SIGCHI Conference on Human Factors in Computing Systems}} (Atlanta, Georgia, USA) \emph{(\bibinfo{series}{CHI '10})}. \bibinfo{publisher}{Association for Computing Machinery}, \bibinfo{address}{New York, NY, USA}, \bibinfo{pages}{567–576}.
\newblock
\showISBNx{9781605589299}
\urldef\tempurl%
\url{https://doi.org/10.1145/1753326.1753410}
\showDOI{\tempurl}


\bibitem[Mirowski et~al\mbox{.}(2023)]%
        {Mirowski23}
\bibfield{author}{\bibinfo{person}{Piotr Mirowski}, \bibinfo{person}{Kory~W. Mathewson}, \bibinfo{person}{Jaylen Pittman}, {and} \bibinfo{person}{Richard Evans}.} \bibinfo{year}{2023}\natexlab{}.
\newblock \showarticletitle{Co-Writing Screenplays and Theatre Scripts with Language Models: Evaluation by Industry Professionals}. In \bibinfo{booktitle}{\emph{Proceedings of the 2023 CHI Conference on Human Factors in Computing Systems}} (Hamburg, Germany) \emph{(\bibinfo{series}{CHI '23})}. \bibinfo{publisher}{Association for Computing Machinery}, \bibinfo{address}{New York, NY, USA}, Article \bibinfo{articleno}{355}, \bibinfo{numpages}{34}~pages.
\newblock
\showISBNx{9781450394215}
\urldef\tempurl%
\url{https://doi.org/10.1145/3544548.3581225}
\showDOI{\tempurl}


\bibitem[Norton et~al\mbox{.}(2012)]%
        {norton2012ikea}
\bibfield{author}{\bibinfo{person}{Michael~I. Norton}, \bibinfo{person}{Daniel Mochon}, {and} \bibinfo{person}{Dan Ariely}.} \bibinfo{year}{2012}\natexlab{}.
\newblock \showarticletitle{The IKEA effect: When labor leads to love}.
\newblock \bibinfo{journal}{\emph{Journal of Consumer Psychology}} \bibinfo{volume}{22}, \bibinfo{number}{3} (\bibinfo{year}{2012}), \bibinfo{pages}{453--460}.
\newblock
\urldef\tempurl%
\url{https://doi.org/10.1016/j.jcps.2011.08.002}
\showDOI{\tempurl}


\bibitem[NovelCraft(2024)]%
        {novelcraft}
\bibfield{author}{\bibinfo{person}{NovelCraft}.} \bibinfo{year}{2024}\natexlab{}.
\newblock \bibinfo{title}{NovelCraft | AI-powered Brainstorming For Writers And Creatives}.
\newblock \bibinfo{howpublished}{\url{https://www.novelcraft.net/}}.
\newblock
\newblock
\shownote{Retrieved: 2024-05-30}.


\bibitem[Novelcrafter(2024)]%
        {novelcrafter}
\bibfield{author}{\bibinfo{person}{Novelcrafter}.} \bibinfo{year}{2024}\natexlab{}.
\newblock \bibinfo{title}{novelcrafter - Your Novel Writing Toolbox}.
\newblock \bibinfo{howpublished}{\url{https://www.novelcrafter.com/}}.
\newblock
\newblock
\shownote{Retrieved: 2024-05-30}.


\bibitem[Papachristos et~al\mbox{.}(2021)]%
        {Papachristos21}
\bibfield{author}{\bibinfo{person}{Eleftherios Papachristos}, \bibinfo{person}{Patrick Skov~Johansen}, \bibinfo{person}{Rune M\o{}berg~Jacobsen}, \bibinfo{person}{Lukas Bj\o{}rn Leer~Bysted}, {and} \bibinfo{person}{Mikael~B. Skov}.} \bibinfo{year}{2021}\natexlab{}.
\newblock \showarticletitle{How do People Perceive the Role of AI in Human-AI Collaboration to Solve Everyday Tasks?}. In \bibinfo{booktitle}{\emph{CHI Greece 2021: 1st International Conference of the ACM Greek SIGCHI Chapter}} (Online (Athens, Greece), Greece) \emph{(\bibinfo{series}{CHI Greece 2021})}. \bibinfo{publisher}{Association for Computing Machinery}, \bibinfo{address}{New York, NY, USA}, Article \bibinfo{articleno}{10}, \bibinfo{numpages}{6}~pages.
\newblock
\showISBNx{9781450385787}
\urldef\tempurl%
\url{https://doi.org/10.1145/3489410.3489420}
\showDOI{\tempurl}


\bibitem[Pierce et~al\mbox{.}(2003)]%
        {pierce2003state}
\bibfield{author}{\bibinfo{person}{Jon~L. Pierce}, \bibinfo{person}{Tatiana Kostova}, {and} \bibinfo{person}{Kurt~T. Dirks}.} \bibinfo{year}{2003}\natexlab{}.
\newblock \showarticletitle{The State of Psychological Ownership: Integrating and Extending a Century of Research}.
\newblock \bibinfo{journal}{\emph{Review of General Psychology}} \bibinfo{volume}{7}, \bibinfo{number}{1} (\bibinfo{year}{2003}), \bibinfo{pages}{84--107}.
\newblock
\urldef\tempurl%
\url{https://doi.org/10.1037/1089-2680.7.1.84}
\showDOI{\tempurl}


\bibitem[Qin et~al\mbox{.}(2024)]%
        {qinCharacterMeetSupportingCreative2024}
\bibfield{author}{\bibinfo{person}{Hua~Xuan Qin}, \bibinfo{person}{Shan Jin}, \bibinfo{person}{Ze Gao}, \bibinfo{person}{Mingming Fan}, {and} \bibinfo{person}{Pan Hui}.} \bibinfo{year}{2024}\natexlab{}.
\newblock \showarticletitle{{{CharacterMeet}}: {{Supporting Creative Writers}}' {{Entire Story Character Construction Processes Through Conversation}} with {{LLM-Powered Chatbot Avatars}}}. In \bibinfo{booktitle}{\emph{Proceedings of the {{CHI Conference}} on {{Human Factors}} in {{Computing Systems}}}}. \bibinfo{publisher}{ACM}, \bibinfo{address}{Honolulu HI USA}, \bibinfo{pages}{1--19}.
\newblock
\showISBNx{9798400703300}
\urldef\tempurl%
\url{https://doi.org/10.1145/3613904.3642105}
\showDOI{\tempurl}


\bibitem[Quillbot(2024)]%
        {quillbot}
\bibfield{author}{\bibinfo{person}{Quillbot}.} \bibinfo{year}{2024}\natexlab{}.
\newblock \bibinfo{title}{Quillbot: Make Writing Painless}.
\newblock \bibinfo{howpublished}{\url{https://quillbot.com/}}.
\newblock
\newblock
\shownote{Retrieved: 2024-05-31}.


\bibitem[Reza et~al\mbox{.}(2025)]%
        {reza2025co}
\bibfield{author}{\bibinfo{person}{Mohi Reza}, \bibinfo{person}{Jeb Thomas-Mitchell}, \bibinfo{person}{Peter Dushniku}, \bibinfo{person}{Nathan Laundry}, \bibinfo{person}{Joseph~Jay Williams}, {and} \bibinfo{person}{Anastasia Kuzminykh}.} \bibinfo{year}{2025}\natexlab{}.
\newblock \showarticletitle{Co-writing with AI, on human terms: Aligning research with user demands across the writing process}.
\newblock \bibinfo{journal}{\emph{arXiv preprint arXiv:2504.12488}} (\bibinfo{year}{2025}).
\newblock


\bibitem[Roemmele(2016)]%
        {roemmele_writing_2016}
\bibfield{author}{\bibinfo{person}{Melissa Roemmele}.} \bibinfo{year}{2016}\natexlab{}.
\newblock \showarticletitle{Writing {Stories} with {Help} from {Recurrent} {Neural} {Networks}}.
\newblock \bibinfo{journal}{\emph{Proceedings of the AAAI Conference on Artificial Intelligence}} \bibinfo{volume}{30}, \bibinfo{number}{1} (\bibinfo{year}{2016}).
\newblock
\urldef\tempurl%
\url{https://ojs.aaai.org/index.php/AAAI/article/view/9810}
\showURL{%
\tempurl}


\bibitem[Roemmele and Gordon(2015)]%
        {roemmele_creative_2015}
\bibfield{author}{\bibinfo{person}{Melissa Roemmele} {and} \bibinfo{person}{Andrew~S. Gordon}.} \bibinfo{year}{2015}\natexlab{}.
\newblock \showarticletitle{Creative {Help}: {A} {Story} {Writing} {Assistant}}.
\newblock In \bibinfo{booktitle}{\emph{Interactive {Storytelling}}}, \bibfield{editor}{\bibinfo{person}{Henrik Schoenau-Fog}, \bibinfo{person}{Luis~Emilio Bruni}, \bibinfo{person}{Sandy Louchart}, {and} \bibinfo{person}{Sarune Baceviciute}} (Eds.). Vol.~\bibinfo{volume}{9445}. \bibinfo{publisher}{Springer International Publishing}, \bibinfo{address}{Cham}, \bibinfo{pages}{81--92}.
\newblock
\showISBNx{978-3-319-27035-7 978-3-319-27036-4}
\urldef\tempurl%
\url{https://doi.org/10.1007/978-3-319-27036-4_8}
\showDOI{\tempurl}
\newblock
\shownote{Series Title: Lecture Notes in Computer Science}.


\bibitem[Sharples and y~P{\'e}rez(2022)]%
        {sharples2022story}
\bibfield{author}{\bibinfo{person}{Mike Sharples} {and} \bibinfo{person}{Rafael~P{\'e}rez y P{\'e}rez}.} \bibinfo{year}{2022}\natexlab{}.
\newblock \bibinfo{booktitle}{\emph{Story machines: How computers have become creative writers}}.
\newblock \bibinfo{publisher}{Routledge}.
\newblock


\bibitem[Shelley(1910)]%
        {shelley1910defence}
\bibfield{author}{\bibinfo{person}{Percy~Bysshe Shelley}.} \bibinfo{year}{1910}\natexlab{}.
\newblock \bibinfo{booktitle}{\emph{A defence of poetry}}. Vol.~\bibinfo{volume}{20}.
\newblock \bibinfo{publisher}{TB Mosher}.
\newblock


\bibitem[Shneiderman(1981)]%
        {shneiderman81direct}
\bibfield{author}{\bibinfo{person}{Ben Shneiderman}.} \bibinfo{year}{1981}\natexlab{}.
\newblock \showarticletitle{Direct manipulation: A step beyond programming languages (abstract only)}.
\newblock \bibinfo{journal}{\emph{SIGSOC Bull.}} \bibinfo{volume}{13}, \bibinfo{number}{2–3} (\bibinfo{date}{May} \bibinfo{year}{1981}), \bibinfo{pages}{143}.
\newblock
\showISSN{0163-5794}
\urldef\tempurl%
\url{https://doi.org/10.1145/1015579.810991}
\showDOI{\tempurl}


\bibitem[Singh et~al\mbox{.}(2022)]%
        {singhWhereHideStolen2022}
\bibfield{author}{\bibinfo{person}{Nikhil Singh}, \bibinfo{person}{Guillermo Bernal}, \bibinfo{person}{Daria Savchenko}, {and} \bibinfo{person}{Elena~L. Glassman}.} \bibinfo{year}{2022}\natexlab{}.
\newblock \showarticletitle{Where to {{Hide}} a {{Stolen Elephant}}: {{Leaps}} in {{Creative Writing}} with {{Multimodal Machine Intelligence}}}.
\newblock \bibinfo{journal}{\emph{ACM Transactions on Computer-Human Interaction}} (\bibinfo{date}{Feb.} \bibinfo{year}{2022}), \bibinfo{pages}{3511599}.
\newblock
\showISSN{1073-0516, 1557-7325}
\urldef\tempurl%
\url{https://doi.org/10.1145/3511599}
\showDOI{\tempurl}


\bibitem[Spittal et~al\mbox{.}(2002)]%
        {Spittal2002}
\bibfield{author}{\bibinfo{person}{M.~J. Spittal}, \bibinfo{person}{R.~J. Siegert}, \bibinfo{person}{J.~L. McClure}, {and} \bibinfo{person}{F.~H. Walkey}.} \bibinfo{year}{2002}\natexlab{}.
\newblock \showarticletitle{The Spheres of Control scale: the identification of a clear replicable three-factor structure}.
\newblock \bibinfo{journal}{\emph{Personality and Individual Differences}} \bibinfo{volume}{32}, \bibinfo{number}{5} (\bibinfo{year}{2002}), \bibinfo{pages}{845--859}.
\newblock
\urldef\tempurl%
\url{https://doi.org/10.1016/S0191-8869(01)00010-1}
\showDOI{\tempurl}


\bibitem[Sudowrite(2024)]%
        {sudowrite}
\bibfield{author}{\bibinfo{person}{Sudowrite}.} \bibinfo{year}{2024}\natexlab{}.
\newblock \bibinfo{title}{Sudowrite - Best AI Writing Partner for Fiction}.
\newblock \bibinfo{howpublished}{\url{https://www.sudowrite.com/}}.
\newblock
\newblock
\shownote{Retrieved: 2024-05-30}.


\bibitem[Svanaes and Verplank(2000)]%
        {Svanaes00}
\bibfield{author}{\bibinfo{person}{Dag Svanaes} {and} \bibinfo{person}{William Verplank}.} \bibinfo{year}{2000}\natexlab{}.
\newblock \showarticletitle{In search of metaphors for tangible user intefaces}. In \bibinfo{booktitle}{\emph{Proceedings of DARE 2000 on Designing Augmented Reality Environments}} (Elsinore, Denmark) \emph{(\bibinfo{series}{DARE '00})}. \bibinfo{publisher}{Association for Computing Machinery}, \bibinfo{address}{New York, NY, USA}, \bibinfo{pages}{121–129}.
\newblock
\showISBNx{9781450373265}
\urldef\tempurl%
\url{https://doi.org/10.1145/354666.354679}
\showDOI{\tempurl}


\bibitem[Troshani et~al\mbox{.}(2021)]%
        {Troshani21}
\bibfield{author}{\bibinfo{person}{Indrit Troshani}, \bibinfo{person}{Sally Rao~Hill}, \bibinfo{person}{Claire Sherman}, {and} \bibinfo{person}{Damien Arthur}.} \bibinfo{year}{2021}\natexlab{}.
\newblock \showarticletitle{Do We Trust in AI? Role of Anthropomorphism and Intelligence}.
\newblock \bibinfo{journal}{\emph{The Journal of Computer Information Systems}} \bibinfo{volume}{61}, \bibinfo{number}{5} (\bibinfo{year}{2021}), \bibinfo{pages}{481--491}.
\newblock
\urldef\tempurl%
\url{https://doi.org/10.1080/08874417.2020.1788473}
\showDOI{\tempurl}


\bibitem[Tzara(2018)]%
        {tzara2018seven}
\bibfield{author}{\bibinfo{person}{Tristan Tzara}.} \bibinfo{year}{2018}\natexlab{}.
\newblock \bibinfo{booktitle}{\emph{Seven Dada manifestos and lampisteries}}.
\newblock \bibinfo{publisher}{Alma Books}.
\newblock


\bibitem[Wan et~al\mbox{.}(2024)]%
        {wan2024ItFeltHaving}
\bibfield{author}{\bibinfo{person}{Qian Wan}, \bibinfo{person}{Siying Hu}, \bibinfo{person}{Yu Zhang}, \bibinfo{person}{Piaohong Wang}, \bibinfo{person}{Bo Wen}, {and} \bibinfo{person}{Zhicong Lu}.} \bibinfo{year}{2024}\natexlab{}.
\newblock \showarticletitle{"{{It Felt Like Having}} a {{Second Mind}}": {{Investigating Human-AI Co-creativity}} in {{Prewriting}} with {{Large Language Models}}}.
\newblock \bibinfo{journal}{\emph{Proceedings of the ACM on Human-Computer Interaction}} \bibinfo{volume}{8}, \bibinfo{number}{CSCW1} (\bibinfo{date}{April} \bibinfo{year}{2024}), \bibinfo{pages}{1--26}.
\newblock
\showISSN{2573-0142}
\urldef\tempurl%
\url{https://doi.org/10.1145/3637361}
\showDOI{\tempurl}


\bibitem[Wasi et~al\mbox{.}(2024)]%
        {wasi2024ink}
\bibfield{author}{\bibinfo{person}{Azmine~Toushik Wasi}, \bibinfo{person}{Raima Islam}, {and} \bibinfo{person}{Mst~Rafia Islam}.} \bibinfo{year}{2024}\natexlab{}.
\newblock \showarticletitle{Ink and Individuality: Crafting a Personalised Narrative in the Age of LLMs}.
\newblock \bibinfo{journal}{\emph{arXiv preprint arXiv:2404.00026}} (\bibinfo{year}{2024}).
\newblock


\bibitem[Whalen(2023)]%
        {whalen2023any}
\bibfield{author}{\bibinfo{person}{Zach Whalen}.} \bibinfo{year}{2023}\natexlab{}.
\newblock \showarticletitle{Any Means Necessary to Refuse Erasure by Algorithm: Lillian-Yvonne Bertram's Travesty Generator.}
\newblock \bibinfo{journal}{\emph{DHQ: Digital Humanities Quarterly}} \bibinfo{volume}{17}, \bibinfo{number}{2} (\bibinfo{year}{2023}).
\newblock


\bibitem[Writesonic(2024)]%
        {writesonic}
\bibfield{author}{\bibinfo{person}{Writesonic}.} \bibinfo{year}{2024}\natexlab{}.
\newblock \bibinfo{title}{Writesonic - AI Content Writer, AI SEO Toolkit \& AI Chatbots}.
\newblock \bibinfo{howpublished}{\url{https://writesonic.com/}}.
\newblock
\newblock
\shownote{Retrieved: 2024-05-31}.


\bibitem[Yuan et~al\mbox{.}(2022)]%
        {yuanWordcraftStoryWriting2022}
\bibfield{author}{\bibinfo{person}{Ann Yuan}, \bibinfo{person}{Andy Coenen}, \bibinfo{person}{Emily Reif}, {and} \bibinfo{person}{Daphne Ippolito}.} \bibinfo{year}{2022}\natexlab{}.
\newblock \showarticletitle{Wordcraft: {{Story Writing With Large Language Models}}}. In \bibinfo{booktitle}{\emph{27th {{International Conference}} on {{Intelligent User Interfaces}}}}. \bibinfo{publisher}{ACM}, \bibinfo{address}{Helsinki Finland}, \bibinfo{pages}{841--852}.
\newblock
\showISBNx{978-1-4503-9144-3}
\urldef\tempurl%
\url{https://doi.org/10.1145/3490099.3511105}
\showDOI{\tempurl}


\end{thebibliography}

\newpage
\appendix{}



\section*{Appendix A: Commercial system Summary Statistics Table}
\label{CASST}
\vspace{-10pt}
\begin{table}[H]
\centering
\small
\setlength{\tabcolsep}{4pt}
\begin{adjustbox}{width=\textwidth,totalheight=\textheight,keepaspectratio}
\begin{tabular}{>{\raggedright\arraybackslash}p{2.8cm} >{\raggedright\arraybackslash}p{4cm} *{8}{c}}
\toprule
\multirow{2}{*}{\textbf{Target Audience}} & \multirow{2}{*}{\textbf{systems Reviewed}} & \multicolumn{3}{c}{\textbf{Writing Processes Supported}} & \multicolumn{5}{c}{\textbf{AI Interaction Interfaces}} \\
\cmidrule(lr){3-5} \cmidrule(lr){6-10}
& & \makecell{\textbf{Planning}} & \makecell{\textbf{Translation}} & \makecell{\textbf{Reviewing}} & \makecell{\textbf{Chat}} & \makecell{\textbf{Rich Text}\\\textbf{Editor}} & \makecell{\textbf{Textbox:}\\\textbf{Ideation}} & \makecell{\textbf{Textbox:}\\\textbf{Translation}} & \makecell{\textbf{Persistent}\\\textbf{Outline}} \\
\midrule
\makecell[l]{\textbf{Academic Writing}\\\textbf{(n = 5)}} & Blainy, CollegeEssay AI Essay Writer, PerfectEssayWriter, TexteroAI, Wordvice AI & 4 & 4 & 5 & 1 & 3 & 4 & 5 & 0 \\
\hline
\makecell[l]{\textbf{Creative Writing}\\\textbf{(n = 10)}} & AutoCrit, DreamGen, Fictionary, NovelAI, NovelCraft, Novelcrafter, Novlr, ProWritingAid, Squibler, Sudowrite & 7 & 6 & 8 & 2 & 6 & 4 & 4 & 6 \\
\hline
\makecell[l]{\textbf{Enterprise AI}\\\textbf{Productivity Suite}\\\textbf{(n = 2)}} & Notion, Writer & 2 & 2 & 2 & 2 & 2 & 2 & 2 & 2 \\
\hline
\makecell[l]{\textbf{General Writing}\\\textbf{(n = 4)}} & ChatGPT, Grammarly, Gemini in Docs, Quillbot & 4 & 4 & 4 & 1 & 2 & 4 & 4 & 1 \\
\hline
\makecell[l]{\textbf{Literacy Support}\\\textbf{(n = 2)}} & CoWriter, Read\&Write & 2 & 2 & 2 & 0 & 2 & 0 & 0 & 0 \\
\hline
\makecell[l]{\textbf{Social Media,}\\\textbf{Marketing, and SEO}\\\textbf{(n = 22)}} & Anyword, Article Forge, Copy.AI, Frase, Helloscribe, HyperWrite, Jasper AI, Junia, Koala, OwlyWriter, Peppertype, Postwise, Protagoras.app, Rytr, Semrush, Simplified, Surfer, Texta, WordAssistant, WordHero, Wordtune, Writesonic & 21 & 22 & 16 & 9 & 16 & 22 & 22 & 9 \\
\bottomrule
\end{tabular}
\end{adjustbox}
\caption{Initial review of AI co-writing systems, showing coding by target audience, writing processes supported by interfaces within that system, and different interaction interfaces}
\label{fig:designreference}
\end{table}

\section*{Appendix B: Interview Protocol for Interview Study}
\label{sec:interview_qs}
\subsection*{Section 1: Introduction}

\begin{itemize}
    \item Explain study goals and introduce interviewer
    \item Discuss participant's process and how system presentation affects them
    \item Inquire about impact of system's look and feel on process and emotions
    \item Request permission to record
\end{itemize}

\subsection*{Section 2: Background}

\begin{enumerate}
    \item Please describe your writing process.
    \item Discuss your relationship with AI and AI writing tools.
\end{enumerate}
\textit{Participants use the AI-writing platform.}

\subsection*{Section 3: General Impressions of AI-Infused Writing System (35 minutes)}

\textit{Participants use one version of the AI platform for 15 minutes.}

\begin{enumerate}
    \item What are your initial impressions of the system?
    \item How, if at all, did your writing process change when using the AI system?
    \item Could you envision yourself or other writers incorporating this system into regular writing practice? Why or why not?
    \item Did you encounter any difficulties while using the system?
    \item How did you decide when to request a generation?
\end{enumerate}

\subsection*{Section 4: Follow-Up Questions on Specific Aspects (20 minutes)}

\textit{Focus on how interaction with the system and its presented metaphor influenced participants' feelings. Encourage references to the specific system used.}

\begin{enumerate}
    \item (Agency) How did the system support or interfere with your control over the writing process?
    \item (Agency) To what extent did you feel the AI system influenced your writing direction?
    \item (Ownership) Does the final product represent your unique style and voice? Why or why not?
    \item (Ownership) Do you feel the final product was co-authored with the AI? Why or why not?
\end{enumerate}

\subsection*{Section 5: Follow-Up Questions on Specific Correlations (20 minutes)}

\begin{enumerate}
    \item Do you feel you collaborated with the AI on this piece? a. How would controlling output length affect your perception of collaboration? b. How would less similar text (e.g., more ornate language) impact your view of collaboration? c. In what ways did the generated text surprise you? How did this affect your sense of control? d. Do you consider the piece co-authored? Who is the primary author?
    \item How would a less familiar writing style or voice in the system affect your sense of control during the interaction?
\end{enumerate}

\end{document}